\documentclass[english,11pt,reqno]{article}

\usepackage[utf8]{inputenc}
\usepackage[english]{babel}

\usepackage{mathrsfs}
\usepackage{amsmath,amsfonts,amssymb}
\usepackage{bbm}
\usepackage{amsthm}

\usepackage{import}

\usepackage{enumerate}

\usepackage{fullpage}

\usepackage{color}

\usepackage{natbib}

\usepackage{multirow}

\usepackage{epsfig}

\usepackage{graphicx}
\usepackage{subfigure}
\usepackage{psfrag}

\usepackage{mathtools}
\mathtoolsset{showonlyrefs}

\newcommand{\addHap}{\ensuremath{\mathfrak{h}}}
\newcommand{\addHapRV}{\ensuremath{\mathfrak{H}}}
\newcommand{\nDemeCfg}[1]{\mathbf{n}_{#1}}

\newcommand{\nCfg}{\ensuremath{\mathbf{n}}}
\newcommand{\nDeme}[1]{\ensuremath{\mathbf{n}_{#1}}}
\newcommand{\nCmp}[2]{\ensuremath{n_{#1,#2}}}
\newcommand{\nSize}{\ensuremath{n}}
\newcommand{\nSizeDeme}[1]{\ensuremath{n_{#1}}}

\newcommand{\sngConfig}[1]{\ensuremath{\mathbf{e}_{#1}}}

\newcommand{\numLoci}{\ensuremath{k}}
\newcommand{\locSet}{\ensuremath{L}}
\newcommand{\loc}{\ensuremath{\ell}}
\newcommand{\locusIdx}{\loc}
\newcommand{\allele}{\ensuremath{a}}
\newcommand{\alleleSet}[1]{\ensuremath{E_{#1}}}
\newcommand{\relDemeSize}[1]{\ensuremath{\kappa_{#1}}}

\newcommand{\hapSpace}{\mathcal{H}}
\newcommand{\hap}{h}
\newcommand{\hapRec}[3]{\mathcal{R}_{#1}(#2,#3)}
\newcommand{\hapSub}[3]{\mathcal{S}_{#1}^{#2}{(#3)}}

\newcommand{\integral}[3]{\ensuremath{\textrm{I}_{#2}^{#3}(#1)}}

\newcommand{\genTot}{\mathscr{L}}
\newcommand{\Ex}{\mathbb{E}}
\newcommand{\ExAppr}{\hat{\mathbb{E}}}
\newcommand{\sampProb}{q}
\newcommand{\sampProbAppr}{\hat{q}}
\newcommand{\genPart}[2]{\mathcal{L}_{#1,#2}}
\newcommand{\stateSpace}{\mathbf{x}}
\newcommand{\stateSpaceRV}{\mathbf{X}}
\newcommand{\stateSpaceCmp}[2]{x_{#1,#2}}
\newcommand{\stateSpaceRVCmp}[2]{X_{#1,#2}}

\newcommand{\addHapCfg}{\ensuremath{\mathbf{c}}}
\newcommand{\addHapCmp}[2]{\ensuremath{c_{#1,#2}}}
\newcommand{\addHapSizeDeme}[1]{\ensuremath{c_{#1}}}
\newcommand{\numAddHap}{\ensuremath{c}}      
\newcommand{\condAnc}[1]{\ensuremath{\mathcal{C}_{#1}}}
\newcommand{\trueAnc}[1]{\mathcal{A}_{#1}}
\newcommand{\trunkAnc}[1]{\ensuremath{\mathcal{A}_0(#1)}}

\newcommand{\mutRate}[1]{\ensuremath{\theta_{#1}}}
\newcommand{\mutRateSingle}{\ensuremath{\theta}}
\newcommand{\rawMutMatrix}{\mathbf{P}}
\newcommand{\mutMatrix}[1]{\rawMutMatrix^{(#1)}}

\newcommand{\recRate}[1]{\ensuremath{\rho_{#1}}}
\newcommand{\recRateSingle}{\ensuremath{\rho}}
\newcommand{\breakTime}{\ensuremath{t_b}}
\newcommand{\breakPoint}{\ensuremath{b}}
\newcommand{\breakPointSet}{\ensuremath{B}}

\newcommand{\demeSpace}{\Gamma}
\newcommand{\numDemes}{\ensuremath{g}}
\newcommand{\deme}{\gamma}

\newcommand{\migRate}[2]{m_{#1#2}}
\newcommand{\migRateDeme}[1]{m_{#1}}
\newcommand{\migRateSingle}{m}
\newcommand{\migProb}[2]{v_{#1#2}}

\newcommand{\migMatrix}{\ensuremath{M}}
\newcommand{\breakDemeRV}[1]{\ensuremath{G^B_{#1}}}
\newcommand{\breakDeme}{\ensuremath{\gamma}}
\newcommand{\addDeme}{\ensuremath{\alpha}}
\newcommand{\nMatrix}{\ensuremath{Z}}

\newcommand{\absTimeRV}[1]{\ensuremath{T^A_{#1}}}
\newcommand{\absTime}[1]{\ensuremath{t_{#1}}}
\newcommand{\absHapRV}[1]{\ensuremath{H^A_{#1}}}
\newcommand{\absHap}[1]{\ensuremath{h_{#1}}}
\newcommand{\absDemeRV}[1]{\ensuremath{G^A_{#1}}}
\newcommand{\absDeme}[1]{\ensuremath{\omega_{#1}}}
\newcommand{\absMatrix}{\ensuremath{A}}
\newcommand{\absState}[1]{\ensuremath{a_{#1}}}

\newcommand{\hState}[1]{\ensuremath{s_{#1}}}
\newcommand{\iDensity}{\ensuremath{\zeta}}
\newcommand{\transDensity}{\ensuremath{\phi}}
\newcommand{\emDensity}{\ensuremath{\xi}}
\newcommand{\fwDensAbsOnly}{\ensuremath{f}}
\newcommand{\hStateSpace}{\ensuremath{S}}

\newcommand{\numPart}{\ensuremath{d}}
\newcommand{\partPoint}[1]{\ensuremath{x_{#1}}}
\newcommand{\partInt}[1]{\ensuremath{D_{#1}}}
\renewcommand{\part}{\ensuremath{\mathcal{D}}}
\newcommand{\partIdx}[1]{\ensuremath{i_{#1}}}
\newcommand{\hStateDisc}[1]{\ensuremath{\sigma_{#1}}}
\newcommand{\iDensityDisc}{\ensuremath{\tilde\zeta}}
\newcommand{\transDensityDisc}{\ensuremath{\tilde\phi}}
\newcommand{\emDensityDisc}{\ensuremath{\tilde\xi}}
\newcommand{\margProb}[2]{\ensuremath{u(#1,#2)}}
\newcommand{\noRecoTrans}[3]{\ensuremath{y_{#3}(#1,#2)}}
\newcommand{\recoTrans}[5]{\ensuremath{z_{#5}(#3,#4|#1,#2)}}
\newcommand{\forwardVar}{\ensuremath{F}}
\newcommand{\hStateSpaceDisc}{\ensuremath{\Sigma}}

\newcommand{\fst}{\ensuremath{\text{F}_\text{ST}}}
\newcommand{\piT}{\ensuremath{\pi_\text{T}}}
\newcommand{\piS}{\ensuremath{\pi_\text{S}}}

\newcommand{\eVal}[1]{\ensuremath{\lambda_{#1}}}
\newcommand{\eVec}[1]{\ensuremath{v_{#1}}}
\newcommand{\doubleU}[1]{\ensuremath{w_{#1}}}
\newcommand{\vMatrix}{\ensuremath{V}}
\newcommand{\evaMatrix}{\ensuremath{\Lambda}}
\newcommand{\eValMut}[1]{\ensuremath{\mu_{#1}}}
\newcommand{\eVecMut}[1]{\ensuremath{q_{#1}}}
\newcommand{\doubleUMut}[1]{\ensuremath{p_{#1}}}
\newcommand{\vMatrixMut}{\ensuremath{Q}}

\newcommand{\pihat}{\hat{\pi}}
\newcommand{\pihatps}{\pihat_{\text{\rm \tiny PS}}}
\newcommand{\pihatls}{\pihat_{\text{\rm \tiny LS}}}

\newcommand{\codeAbsOnly}{\ensuremath{\text{MigSMC-AO}}}

\newcommand{\pihatmig}{\pihat_{\text{\rm \tiny Mig}}}
\newcommand{\pihatmigsmc}{\pihat_{\text{\rm \tiny MigSMC}}}
\newcommand{\pihatmigsmcAO}{\pihat_{\text{\rm \tiny \codeAbsOnly}}}
\newcommand{\pihatmigsmcDisc}{\pihat_{\text{\rm \tiny MigSMC-AOD}}}

\DeclareMathOperator*{\LCL}{\text{LCL}}
\DeclareMathOperator*{\PAC}{\text{PAC}}
\DeclareMathOperator*{\PCL}{\text{PCL}}

\numberwithin{equation}{section}

\theoremstyle{plain}

\newcommand{\sref}[1]{Section~\ref{sec:#1}}
\newcommand{\fref}[1]{Figure~\ref{fig:#1}}

\theoremstyle{definition}

\theoremstyle{remark}
\newtheorem*{remark}{Remark}

 \newcounter{todocounter}
\newcommand{\R}{\ensuremath{\mathbb{R}}}

\renewcommand{\P}{\ensuremath{\mathbb{P}}}
\newcommand{\1}{\ensuremath{\mathbf{I}}}

\begin{document}

\title{A sequentially Markov conditional sampling distribution for structured populations with migration and recombination}

\author{Matthias Steinr\"ucken$^{1,*}$, Joshua S. Paul$^{2,\dagger}$ and Yun S. Song$^{1,2,\ddagger}$}

\footnotetext[1]{Department of Statistics, University of California, Berkeley, CA 94720, USA}
\footnotetext[2]{Computer Science Division, University of California, Berkeley, CA 94720, USA}



\date{}

\maketitle

\begin{abstract}
Conditional sampling distributions (CSDs), sometimes referred to as copying models, underlie numerous practical tools in population genomic analyses. Though an important application that has received much attention is the inference of population structure, the explicit exchange of migrants at specified rates has not hitherto been incorporated into the CSD in a principled framework. Recently, in the case of a single panmictic population, a sequentially Markov CSD has been developed as an accurate, efficient approximation to a principled CSD derived from the diffusion process dual to the coalescent with recombination. In this paper, the sequentially Markov CSD framework is extended to incorporate subdivided population structure, thus providing an efficiently computable CSD that admits a genealogical interpretation related to the structured coalescent with migration and recombination.  As a concrete application, it is demonstrated empirically that the CSD developed here can be employed to yield accurate estimation of a wide range of migration rates.
\renewcommand{\thefootnote}{}\footnote{Email: ${}^*${\tt steinrue@stat.berkeley.edu}, ${}^\dagger${\tt jspaul@cs.berkeley.edu}, ${}^\ddagger${\tt yss@stat.berkeley.edu}}
\end{abstract}

Keywords:
structured coalescent, recombination, migration, conditional sampling distribution, hidden Markov model, sequentially Markov coalescent



\section{Introduction}
\label{sec_introduction}

Under a given population genetic model, the conditional sampling distribution (CSD), also called a \emph{copying model} by some authors, describes the probability that an additionally sampled haplotype is of a certain type, given that a collection of haplotypes has already been observed.   As described below, various applications in population genomics make use of the CSD. Although the CSD is of much importance, no exact closed-form expressions are known in the situations to which it has been applied, and so a number of approximations have been proposed.

Following the seminal work of \citet{Stephens2000} and \citet{Fearnhead2001}, \citet{Li2003} proposed a widely used CSD, denoted $\pihatls$, which models the additionally observed haplotype as an imperfect mosaic of the haplotypes already observed.  The model underlying $\pihatls$ can be cast as a hidden Markov model (HMM), thus admitting efficient implementation.   In their paper, \citeauthor{Li2003} used the CSD $\pihatls$ in a pseudo-likelihood framework to estimate fine-scale recombination rates, and subsequently $\pihatls$ and its extensions have been used in numerous other population genetic applications, including estimating gene-conversion parameters \citep{Gay2007,Yin2009}, and phasing genotype sequence data into haplotype sequence data and imputing missing data \citep{Stephens2005, Li2006, Li2010, Marchini2007, Howie2009}.

Another important application of the CSD that has received much attention is the inference of population structure and demography. \citet{Hellenthal2008} employed $\pihatls$ to model human colonization history as a sequence of founder events and estimated the order of the founding events, as well as the relative contribution of different founding populations during the events. To estimate the splitting time of two populations \citet{Davison2009} modified $\pihatls$ to incorporate the split into the copying model, and used the same pseudo-likelihood framework as \cite{Li2003} to estimate the time of splitting. In a more recent study, \citet{Lawson2012} applied $\pihatls$ to a sample of DNA sequences and used properties of the inferred mosaic pattern to reveal structure in the underlying population.  

To handle admixture, a modification to $\pihatls$ was introduced by \citet{Price2009}, who assumed that the previously observed haplotypes in the CSD are from two distinct ancestral populations (e.g., African and European).  In modeling the mosaic pattern for a haplotype sampled from the admixed population (e.g., African American), it is then assumed more likely that adjacent segments originate from the same ancestral population, rather than from two different ancestral populations.  \citeauthor{Price2009} applied this modified copying model to detect chromosomal segments of distinct ancestry in admixed individuals and estimated admixture fractions in recently admixed populations.  The same model was applied by \citet{Wegmann2011}, who used the inferred ancestry switch-points to estimate relative recombination rates between different populations.

As discussed above, $\pihatls$ is a very useful CSD with a variety of applications, but it was not \emph{derived} from, though was certainly motivated by, principles underlying the coalescent process.  To derive CSDs in a principled way, \citet{DeIorio2004a} introduced a general approximation technique based on the diffusion process dual to the coalescent; this work was first presented in the case of a single locus and a panmictic population, but in a companion paper \citep{DeIorio2004b} the authors applied the method to the case of a subdivided population with migration.
\citet{Griffiths2008} extended the diffusion approximation technique to handle recombination in the special case of two loci with parent-independent mutation at each locus, and \citet{Paul2010} later generalized the framework to an arbitrary number of loci and an arbitrary finite-alleles mutation model.

Although more accurate than the CSDs developed by \citet{Fearnhead2001} and by \citet{Li2003}, the CSD $\pihatps$ proposed by \citet{Paul2010} is not amenable to efficient evaluation.  More precisely, $\pihatps$ can be computed by solving a recursion that becomes intractable for a large number of loci.  However, utilizing ideas related to the sequentially Markov coalescent (SMC) \citep{Wiuf1999,McVean2005,Marjoram2006}, which is a simplified genealogical process that captures the essential features of the full coalescent model with recombination, we \citep{Paul2011} recently developed an approximation to $\pihatps$ that could be cast as an HMM with continuous hidden state space.  Furthermore, upon discretizing this continuous state space, we obtained an accurate approximation with computational efficiency comparable to the CSDs of \citet{Fearnhead2001} and \citet{Li2003}.

In this paper, we extend our previous work on the sequentially Markov CSD to incorporate subdivided population structure with migration. Following \citet{Paul2010}, we describe a genealogical process for an additionally sampled haplotype conditioned on the genealogy of already observed haplotypes. We present a recursion that can be used to compute the probability of the additionally sampled haplotype, but, as in \citet{Paul2010}, solving this recursion is tractable only for a small number of loci.  As in \citet{Paul2011}, we apply the sequentially Markov framework to the conditional genealogical process with migration and recombination, and obtain an accurate approximation that facilitates computation for a large number of loci.  As a concrete application, we demonstrate empirically that our new CSD can be employed in various pseudo-likelihoods to produce accurate estimation of a wide range of migration rates. 

The remainder of this paper is organized as follows: In Section~\ref{sec_background}, we introduce the notation adopted throughout the paper and describe the relevant population genetic model, the coalescent with recombination and migration. We then describe the genealogical interpretation of our CSD in Section~\ref{sec:pihatmig} and introduce several approximations in Section~\ref{sec_approximations} to obtain a CSD for which computation is tractable.  In Section~\ref{sec_estimate_migration_rates}, we demonstrate the applicability of our CSD by employing it to the estimation of migration rates from simulated data.  Finally, we conclude in Section~\ref{sec_discussion} with a discussion of further applications and extensions of the CSD developed herein to estimate demographic parameters in more complex scenarios.

\section{Background} \label{sec_background}
In this section, we briefly describe how migration is integrated into the coalescent with recombination, and recall the CSD $\pihatps$ proposed by \citet{Paul2010}, which we extend to incorporate migration in the following section. We begin by defining some general notation that will be used throughout.

\subsection{Notation} \label{sec:notation}
We consider haplotypes in the finite-sites, finite-alleles setting. Denote the set of loci by $\locSet = \{1,\ldots,\numLoci\}$ and the set of alleles at locus $\loc \in \locSet$ by $\alleleSet{\loc}$; recombination may occur between any consecutive pair of loci, and we denote the set of potential recombination breakpoints by $\breakPointSet = \{(1,2),\ldots,(\numLoci-1,\numLoci)\}$. The space of $\numLoci$-locus haplotypes is denoted by $\hapSpace = \alleleSet{1}\times\cdots\times\alleleSet{\numLoci}$. Given a haplotype $\hap \in \hapSpace$, we denote by $\hap[\loc] \in \alleleSet{\loc}$ the allele at locus $\loc \in \locSet$, and by $\hap[\loc:\loc']$ (for $\loc \leq \loc'$) the partial haplotype $(\hap[\loc],\ldots,\hap[\loc'])$. 
                                  
We consider an island model of population structure with a finite set of demes denoted by $\demeSpace = \{1,\ldots,\numDemes\}$.  At a given time, each individual belongs to a single deme, and the ancestors and descendants of the individual may belong to different demes by means of a migration process, detailed in \sref{coa_rec_mig}.

A \emph{structured sample configuration} of haplotypes is specified by $\nCfg = \big(\nCmp{\deme}{\hap}\big)_{\deme\in\demeSpace, \hap\in\hapSpace}$, where $\nCmp{\deme}{\hap}$ denotes the number of haplotypes of type $\hap$ within deme $\deme$ in the sample. The configuration of haplotypes within deme $\deme\in\demeSpace$ is denoted $\nDemeCfg{\deme}$, and the total number of haplotypes in the deme by $|\nDemeCfg{\deme}| = \nSizeDeme{\deme}$.  The total number of haplotypes in $\nCfg$ is denoted by $\nSize =|\nCfg|=\sum_{\deme\in\demeSpace} \nSizeDeme{\deme}$. Finally, we use $\sngConfig{\deme,\hap}$ to denote the singleton configuration comprising a single haplotype $\hap$ within deme $\deme$.

\subsection{The coalescent with recombination and migration} \label{sec:coa_rec_mig}
The stochastic process underlying our work is the coalescent with recombination and migration \citep{Griffiths1997,Herbots1997}. Consider a structured population with a finite set $\demeSpace$ of demes.  We denote the relative size of deme $\deme\in\demeSpace$ by $\relDemeSize{\deme}$, where $0 \leq \relDemeSize{\deme} \leq 1$ and $\sum_{\deme\in\demeSpace} \relDemeSize{\deme} = 1$. Note that two individuals within a deme find a common ancestor at rate inversely proportional to the relative size of the deme; in the coalescent limit, coalescence within deme $\deme$ occurs at rate $\relDemeSize{\deme}^{-1}$. 

To allow for migration of ancestral lineages between demes, define $\migProb{\deme}{\deme'}$ to be the probability that an individual in deme $\deme$ has a parent in deme $\deme'$. In the coalescent limit, as the population size $N$ tends to infinity, an ancestral lineage in deme $\deme$ migrates, backwards in time, to deme $\deme'$ at rate $\migRate{\deme}{\deme'} / 2$, where $\migRate{\deme}{\deme'} = 4 N \migProb{\deme}{\deme'}$ is the \emph{scaled} migration rate.  Henceforward, we consider a continuous-time Markov migration process with transition rate matrix $\migMatrix = (\migRate{\deme}{\deme'} / 2)_{\deme,\deme' \in \demeSpace}$, where $\migRate{\deme}{\deme} = - \sum_{\deme' \neq \deme} \migRate{\deme}{\deme'}$.  For ease of notation, we define $\migRateDeme{\deme} = \sum_{\deme' \neq \deme} \migRate{\deme}{\deme'}$.
                                                                                                            
An ancestral lineage undergoes mutation at locus $\loc\in\locSet$ at rate $\mutRate{\loc} / 2$, where $\mutRate{\loc}$ is the scaled mutation rate, and according to the stochastic mutation transition matrix $\mutMatrix{\loc}$.  Further, as in the ordinary coalescent with recombination, an ancestral lineage undergoes recombination, backwards in time, at breakpoint $\breakPoint\in\breakPointSet$ at rate $\recRate{\breakPoint}/2$, where $\recRate{\breakPoint}$ is the scaled recombination rate, giving rise to two lineages (in the same deme).

A structured configuration $\nCfg$ with $\nSizeDeme{\deme}$ individuals in each deme $\deme$ can be sampled as follows.  The process starts at present with $\nSizeDeme{\deme}$ untyped lineages in each deme $\deme$, and lineages in each deme $\deme$ evolve backwards in time subject to the following possible events:
\begin{description}
	\item[\hspace{4mm} Mutation:] Each lineage undergoes mutation at locus $\loc \in \locSet$ with rate $\mutRate{\loc} / 2$ according to the mutation transition matrix $\mutMatrix{\loc}$.
	\item[\hspace{4mm} Recombination:] Each lineage undergoes recombination at breakpoint $\breakPoint \in \breakPointSet$ with rate $\recRate{\breakPoint} / 2$.
	\item[\hspace{4mm} Migration:] Each lineage migrates to deme $\deme'$ with rate $\migRate{\deme}{\deme'} / 2$.
	\item[\hspace{4mm} Coalescence:] Each pair of lineages coalesce with rate $\relDemeSize{\deme}^{-1}$. 
\end{description}
When a single lineage remains, the type at each locus $\loc$ is chosen according to the stationary distribution of the mutation matrix $\mutMatrix{\loc}$, and this type is propagated toward the present, producing a realization for the sample $\nCfg$.

\subsection{The CSD $\pihatps$  for a single panmictic population} \label{sec:pihatps}
The approximate CSD $\pihatps$ \citep{Paul2010} for a single panmictic population is described by a genealogical process closely related to the coalescent with recombination.   Suppose that, conditioned on having already observed a haplotype configuration $\nCfg$, we wish to sample $\numAddHap$ additional haplotypes.  As described in \citet{Paul2010}, given the true fully-specified genealogy $\trueAnc{\nCfg}$ for the conditional configuration $\nCfg$, it is possible to sample a \emph{conditional genealogy} $\condAnc{}$ for the $\numAddHap$ additional haplotypes.

The conditional genealogy $\condAnc{}$ comprises the following: mutation, recombination, and coalescence within $\condAnc{}$, occurring at rates given in \sref{coa_rec_mig}; and \emph{absorption} of lineages into the known genealogy $\trueAnc{\nCfg}$, occurring at rate $1$ for each pair. Because the types of the lineages of $\trueAnc{\nCfg}$ are known, the type of an absorbed lineage is determined. Thus, when all lineages of $\condAnc{}$ have been absorbed, the type may be propagated forward, thereby producing a realization for sample configuration $\addHapCfg$ with $|\addHapCfg|=\numAddHap$.

Unfortunately, we do not generally have access to the true genealogy $\trueAnc{\nCfg}$. Making use of the diffusion-generator approximation \citep{DeIorio2004a, DeIorio2004b, Griffiths2008}, \citet{Paul2010} propose the following: \emph{Assume} that $\trueAnc{\nCfg} = \trunkAnc{\nCfg}$, where $\trunkAnc{\nCfg}$ is called the \emph{trunk} genealogy in which lineages do not mutate, recombine, or coalesce with one another, but instead form a non-random ``trunk'' extending infinitely into the past. Note that $\trunkAnc{\nCfg}$ does not have a most recent common ancestor, and for this reason is improper; nonetheless, it remains possible to sample a well-defined conditional genealogy $\condAnc{}$, and thus to generate an additional sample $\addHapCfg$, in much the same way as described above. In particular, lineages within $\condAnc{}$ evolve backwards in time subject to the following events:
\begin{description}
	\item[\hspace{4mm} Mutation:] Each lineage undergoes mutation at locus $\loc \in \locSet$ with rate $\mutRate{\loc}$ according to $\mutMatrix{\loc}$.
	\item[\hspace{4mm} Recombination:] Each lineage undergoes recombination at breakpoint $\breakPoint\in\breakPointSet$ with rate $\recRate{\breakPoint}$
	\item[\hspace{4mm} Coalescence:] Each pair of lineages coalesce with rate $2$.
	\item[\hspace{4mm} Absorption:] Each lineage is absorbed into a lineage of $\trueAnc{\nCfg} = \trunkAnc{\nCfg}$ with rate $1$.
\end{description}
Observe that the rate of absorption is the same as in the case where $\trueAnc{\nCfg}$ is known. The rates for mutation, recombination, and coalescence, on the other hand, are each a factor of two larger than those given in \sref{coa_rec_mig}; intuitively, this adjustment accounts for using the (incorrect) trunk genealogy $\trunkAnc{\nCfg}$, and notably the absence of events therein. Importantly, the CSD $\pihatps$ has been shown to be \emph{correct} for a one-locus model with parent independent mutation \citep{Stephens2000,DeIorio2004a,Paul2010}, a strong argument in favor of the given rate adjustment. The CSD $\pihatps$ is completely characterized by the above genealogical process.

\begin{remark}
	The rates given for the genealogical process governing $\pihatps$ are \emph{double} those given by \citet{Paul2010} and \citet{Paul2011}. Importantly, the genealogical process is \emph{time-homogeneous}, and so for the purposes of computing the conditional sampling probability (CSP) $\pihatps(\addHapCfg | \nCfg)$, this modification has no effect (indeed, any constant multiple of the rates will yield the same CSP). However, we believe that the scaling adopted here admits a natural interpretation of the absorption time as a true coalescence time. For example, consider sampling a single haplotype conditional on a configuration $\nCfg$ with $|\nCfg|=1$; analogous to coalescence of two lines in Kingman's coalescent, absorption in the genealogical process associated with $\pihatps$ occurs at rate 1.
\end{remark}

\section{A new CSD $\pihatmig$ for structured populations with recombination and migration} \label{sec:pihatmig}

We now introduce an approximate CSD $\pihatmig$ by extending the genealogical process of \sref{pihatps} to a general structured population with $|\demeSpace| \geq 1$. Suppose that conditioned on having already observed a structured sample configuration $\nCfg$, we wish to sample $\numAddHap$ additional haplotypes with $\addHapSizeDeme{\deme}$ of them in each deme $\deme$. As before, given the true fully-specified genealogy $\trueAnc{\nCfg}$ for the conditional configuration $\nCfg$, including migration events, it is possible to sample a conditional genealogy $\condAnc{}$ for the $\numAddHap$ additional haplotypes. The conditional genealogy $\condAnc{}$ comprises the events and corresponding rates of \sref{coa_rec_mig}, this time including migration, and the \emph{absorption} of lineages in each deme $\deme$ into lineages of $\trueAnc{\nCfg}$ in deme $\deme$. These latter absorption events occur at rate $\relDemeSize{\deme}^{-1}$.

\begin{figure}
	\centering
	%
	\begin{minipage}[t]{.45\textwidth}\centering
	\subfigure[$\pihatmig$]{
		\label{fig_csd_genealogical}
		\includegraphics[width=1\textwidth]{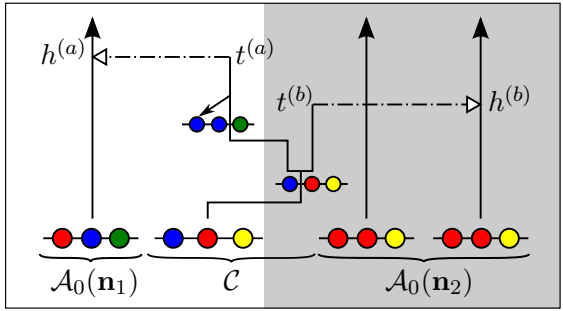}
	}

	\subfigure[$\pihatmigsmc$]{
		\label{fig_csd_smc}
		\includegraphics[width=1\textwidth]{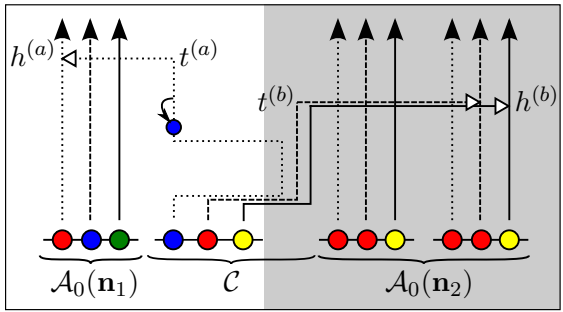}
	}

	\subfigure[$\pihatmigsmcAO$]{
		\label{fig_csd_absorption_only}
		\includegraphics[width=1\textwidth]{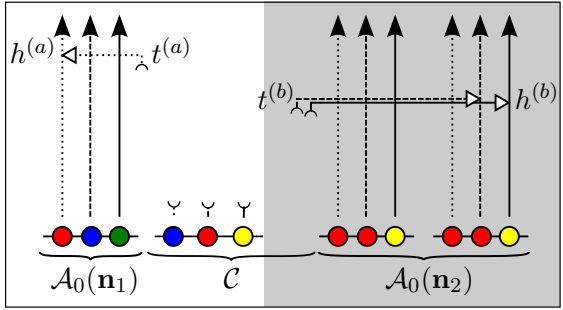}
	}
	\end{minipage}
	\begin{minipage}[t]{.1\textwidth}\phantom{space}\end{minipage}
	\begin{minipage}[t]{.45\textwidth}\centering\vspace{-0.2cm}
	\caption{Illustration of the subsequent approximations to the true conditional sampling distribution. The three loci of each haplotype are each represented by a filled circle, with the color indicating the allelic type at that locus. The trunk genealogies in deme 1 $\trunkAnc{\nDeme{1}}$ and deme 2 $\trunkAnc{\nDeme{2}}$, as well as the conditional genealogy $\condAnc{}$ are indicated. The different demes are indicated by the white and the grey background. Time is represented vertically, with the present (time 0) at the bottom of the illustration. \subref{fig_csd_genealogical} The genealogical interpretation: Mutation events, along with the locus and resulting haplotype, are indicated by small arrows. Recombination events, and the resulting haplotype, are indicated by branching events in $\condAnc{}$. Migration events are indicated by switching to another deme. Absorption events, and the corresponding absorption time ($\absTime{}^{(a)}$ and $\absTime{}^{(b)}$) and haplotype ($\absHap{}^{(a)}$ and $\absHap{}^{(b)}$, respectively), are indicated by dot-dashed horizontal lines. \subref{fig_csd_smc} The corresponding sequential interpretation: The marginal genealogies at the first, second, and third locus are emphasized as dotted, dashed, and solid lines, respectively. Mutation events at each locus, along with resulting allele, are indicated by small arrows. Absorption events at each locus are indicated by horizontal lines. \subref{fig_csd_absorption_only} The corresponding sequential interpretation where just the deme of absorption, the time of absorption, and the absorbing haplotype are recorded. The gap in the ancestral lineages indicates that the marginal conditional genealogy is integrated out.}
	\label{fig_csd}
	\end{minipage}
\end{figure}

In practice, we do not have access to the true genealogy $\trueAnc{\nCfg}$, but the diffusion-generator technique \citep{DeIorio2004a, DeIorio2004b, Griffiths2008, Paul2010} again implies the following approximation: \emph{Assume} that $\trueAnc{\nCfg} = \trunkAnc{\nCfg} = \{\trunkAnc{\nDemeCfg{\deme}}\}_{\deme\in\demeSpace}$, where $\trunkAnc{\nDemeCfg{\deme}}$ is the non-random sub-trunk genealogy associated with deme $\deme$, within which lineages do not mutate, recombine, migrate, or coalesce with one another. As in \sref{pihatps}, given this assumption it remains possible to sample a well-defined conditional genealogy $\condAnc{}$, and thus to generate the additional structured sample $\addHapCfg$. Specifically, lineages within each deme $\deme$ of $\condAnc{}$ evolve backwards in time subject to the following events:
\begin{description}
	\item[\hspace{4mm} Mutation:] Each lineage undergoes mutation at locus $\loc \in \locSet$ with rate $\mutRate{\loc}$ according to the mutation transition matrix $\mutMatrix{\loc}$.
	\item[\hspace{4mm} Recombination:] Each lineage undergoes recombination at breakpoint $\breakPoint\in\breakPointSet$ with rate $\recRate{\breakPoint}$.
	\item[\hspace{4mm} Migration:] Each lineage migrates to deme $\deme'$ with rate $\migRate{\deme}{\deme'}$.
	\item[\hspace{4mm} Coalescence:] Each pair of lineages coalesces with rate $2\relDemeSize{\deme}^{-1}$.
	\item[\hspace{4mm} Absorption:] Each lineage is absorbed into a lineage of $\trunkAnc{\nDemeCfg{\deme}}$ with rate $\relDemeSize{\deme}^{-1}$.
\end{description}
Observe that the rates of mutation, recombination, migration, and coalescence are a factor of two larger than when the true genealogy $\trueAnc{\nCfg}$ is known.  Intuitively, this again accounts for using the (incorrect) trunk genealogy $\trunkAnc{\nCfg}$, and particularly the absence of events therein; see the remark at the end of Section~\ref{sec:pihatps}.  The approximate CSD $\pihatmig$ is completely characterized by this genealogical process. See Figure~\ref{fig_csd_genealogical} for an illustration.

\begin{remark}
For strongly asymmetric migration rates, the approximate CSD $\pihatmig$, and in particular the assumed trunk genealogy $\trunkAnc{\nCfg}$, may be very inaccurate. Consider for example the case of two demes and $\migRate{1}{2} \gg \migRate{2}{1}$. The expected time for an additionally sampled haplotype in deme 2 to be absorbed into the trunk in deme 1 will be very large, since the lineages in the trunk genealogy $\trunkAnc{\nDemeCfg{1}}$ are confined to stay in deme $1$. In case of the true genealogy $\trueAnc{\nCfg}$, however, one would expect the lineages of the haplotypes in the observed configuration in deme 1 to cross over to deme 2 quickly and coalesce more recently with the additional lineage.
\end{remark}

We now consider computing the CSP $\pihatmig(\addHapCfg | \nCfg)$.  It is possible to derive the following result directly using the diffusion-generator approximation, but we defer this work to Appendix~\ref{sec:app_diff_apprx}.  Below, we obtain the result through the genealogical process detailed above; using typical forward-backward genealogical arguments in coalescent theory, we deduce that $\pihatmig(\addHapCfg | \nCfg)$ satisfies the following equation:
\begin{align}\label{eq_recursion}
		\pihatmig(\addHapCfg | \nCfg) 
			& = \frac {1} {\mathcal{N}} \cdot \sum_{\deme \in \demeSpace, \atop \hap \in \hapSpace} \addHapCmp{\deme}{\hap} \bigg[ \Big(\nCmp{\deme}{\hap} + \addHapCmp{\gamma}{\hap} - 1\Big) \relDemeSize{\gamma}^{-1} \pihatmig (\addHapCfg - \sngConfig{\deme,\hap} | \nCfg) \nonumber \\
			&\qquad + \sum_{\loc\in\locSet} \mutRate{\loc} \sum_{\allele \in \alleleSet{\loc}} \mutMatrix{\loc}_{\allele,\hap[\loc]} \pihatmig(\addHapCfg - \sngConfig{\deme,\hap} + \sngConfig{\deme,\hapSub{\loc}{\allele}{\hap}} | \nCfg) \nonumber \\
			&\qquad + \sum_{\breakPoint \in \breakPointSet} \recRate{\breakPoint} \sum_{\hap' \in \hapSpace} \pihatmig(\addHapCfg - \sngConfig{\deme,\hap} + \sngConfig{\deme,\hapRec{\breakPoint}{\hap}{\hap'}} + \sngConfig{\deme,\hapRec{\breakPoint}{\hap'}{\hap}} | \nCfg) \nonumber \\
			&\qquad + \sum_{\deme' \neq \deme} \migRate{\deme}{\deme'} \pihatmig (\addHapCfg - \sngConfig{\deme,\hap} + \sngConfig{\deme',\hap} | \nCfg) \bigg],
\end{align}
where $\hapSub{\loc}{\allele}{\hap}$ denotes the haplotype obtained by substituting the allele at locus $\loc$ of $\hap$ with allele $\allele$, and $\hapRec{\breakPoint}{\hap}{\hap'}$ denotes the haplotype obtained, via recombination about breakpoint $\breakPoint = (\loc,\loc+1)$, by joining the (partial) haplotypes $\hap[1:\loc]$ and $\hap'[\loc+1,\numLoci]$. The first term on the right hand side of this equation corresponds to coalescence and absorption of haplotype $\hap$ in deme $\deme$, and the subsequent terms correspond to mutation, recombination, and migration, respectively. The normalizing constant $\mathcal{N}$ is given by
\begin{equation*}
	\mathcal{N} = \sum_{\deme \in \demeSpace} \addHapSizeDeme{\deme} \Bigg[ (\nSizeDeme{\deme} + \addHapSizeDeme{\deme} - 1 ) \relDemeSize{\deme}^{-1} + \sum_{\loc\in \locSet} \mutRate{\loc} + \sum_{\breakPoint \in \breakPointSet} \recRate{\breakPoint} + \sum_{\deme'\neq\deme} \migRate{\deme}{\deme'} \Bigg].
\end{equation*}
Equation \eqref{eq_recursion} is for the ``full'' (conditional) genealogical process, and, because of the recombination terms, it cannot be directly computed by solving a set of linear equations.  However, as in \citet{Paul2010}, it is possible to derive a ``reduced'' recursion related to \eqref{eq_recursion} that can be computed by solving a finite set of linear equations. Unfortunately, the number of variables in the set of equations grows super-exponentially with both the number of loci and the number of haplotypes in the sample configuration $\nCfg$, making it computationally intractable for all but the smallest problems.   In the following section, we propose accurate approximations that allow for efficient computation.

\section{An efficiently computable CSD as an approximation of $\pihatmig$}
\label{sec_approximations}
As described above, the recursion for $\pihatmig(\addHapCfg | \nCfg)$ becomes computationally intractable for even modest datasets.  In what follows, we adopt a set of approximations to obtain a CSD that admits efficient implementation, while retaining the accuracy of $\pihatmig$.

\subsection{The CSD $\pihatmigsmc$: Sequentially Markov approximation of $\condAnc{}$}
\label{sec_smc}

We follow \citet{Paul2011} and use ideas underlying the SMC \citep{Wiuf1999,McVean2005,Marjoram2006} to approximate $\pihatmig$. Briefly, observe that a given conditional genealogy induces a \emph{marginal conditional genealogy} (MCG) at each locus, where each MCG comprises a series of mutation and migration events, and the eventual absorption into a lineage of the sub-trunk in a certain deme. See Figure~\ref{fig_csd_smc} for an illustration. The key insight, initially provided by \citet{Wiuf1999}, is that we can generate the conditional genealogy as a \emph{sequence} of MCGs, rather than backwards in time. Moreover, though the sequence is not formally Markov, it is well approximated \citep{McVean2005,Marjoram2006,Paul2011} by a Markov process using a two-locus transition density. Applying this approximation to $\pihatmig$ yields the \emph{sequentially Markov} CSD $\pihatmigsmc$. For ease of exposition, we restrict attention to the case of sampling a single additional haplotype, denoted $\addHap$, but the ideas generalize, in principle, to sampling two or more additional haplotypes. 


Since mutations can be superimposed onto the conditional genealogy, we first consider generating a sequence of MCGs without mutations according to a Markov process.
The genealogical process underlying $\pihatmig$ yields the following sampling procedure for the MCG at an arbitrary locus: The ancestral lineage of the additionally sampled haplotype initially resides in deme $\addDeme{}$, where the additional haplotype is sampled, and proceeds backwards in time, subject to the migration process, until being absorbed into a lineage of the sub-trunk $\trunkAnc{\nDemeCfg{\deme}}$ within the current deme $\deme$. The associated marginal distribution is used as the initial distribution at the first locus.

\begin{figure}
	\centering
	
	%
	\includegraphics[width=0.5\textwidth]{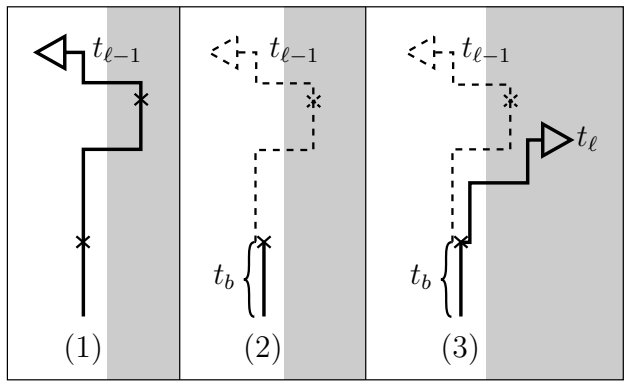}

	\caption{The transition density from locus $\locusIdx-1$ to locus $\locusIdx$ in the model underlying $\pihatmigsmc$ is illustrated. The white and the grey background symbolize the two different demes that the ancestral lineage can reside in. (1) A Poisson number of recombination events is placed uniformly onto the marginal conditional genealogy at locus $\locusIdx-1$. (2) If the time $\breakTime$ of the most recent recombination event is more recent than the time of absorption $\absTime{\locusIdx-1}$, then the marginal conditional genealogy up to this time is copied to locus $\locusIdx$. (3) The ancestral lineage at locus $\locusIdx$ evolves according to migration until it is absorbed at time $\absTime{\locusIdx}$ into the trunk in some deme.}
	\label{fig_transition}
\end{figure}

Conditional on the marginal genealogy at locus $\locusIdx-1$, the marginal genealogy at locus $\locusIdx$ can be sampled by first placing recombination events onto the MCG at locus $\locusIdx-1$ according to a Poisson process with rate $\recRate{(\locusIdx-1,\locusIdx)}$. If no recombination occurs, the marginal genealogy at locus $\loc$ is identical to the one at locus $\loc - 1$. If recombination does occur, the MCG at locus $\locusIdx$ is identical to the MCG at locus $\loc - 1$ up to the time $\breakTime$ of the most recent recombination event. At this point, the lineage resides in the same deme in which the ancestral lineage at locus $\locusIdx-1$ resided at the time of the recombination event, and, independently of the lineage at locus $\loc -1$, proceeds backwards in time, subject to the migration process, until being absorbed into a lineage of the sub-trunk $\trunkAnc{\nDemeCfg{\deme}}$ within the current deme $\deme$. Figure~\ref{fig_transition} illustrates this transition mechanism for the Markov process.

Conditional on the MCG at locus $\loc$, mutations are superimposed onto the MCG according to a Poisson process with rate $\mutRate{\locusIdx}$. The MCG is absorbed into a trunk lineage corresponding to some haplotype $\hap$, thereby specifying an ``ancestral'' allele $\hap[\loc]$. This allele is then propagated to the present according to the mutations and the mutation transition matrix $\mutMatrix{\locusIdx}$, thereby generating an allele at locus $\loc$ of the additional haplotype. We refer to the associated distribution of alleles as the emission distribution.

It is possible to write down explicit expressions for the initial, transition, and emission distributions for $\pihatmigsmc$. However, as the state space for the MCG at each locus includes the entire migrational history, an efficient algorithm for computing the CSP is not known. In the next subsections, we introduce further approximations to this model in order to admit an efficient implementation.

Although we do not prove it here, we note that, analogous to \citet{Paul2011}, the sequentially Markov version of the CSD can be obtained from the genealogical process introduced in Section~\ref{sec:pihatmig} by prohibiting coalescence events in the conditional genealogy between lineages not ancestral to any overlapping parts of the additionally sampled haplotypes. In the case of sampling one additional haplotype, this corresponds to prohibiting all coalescence events in the conditional genealogy. This observation allows one to write down a recursive formula to compute probabilities under $\pihatmigsmc$, but this again does not lead to an efficient implementation.

\subsection{The CSD  $\pihatmigsmcAO$: Keeping track of the absorption time only}
\label{sec_absorption_time_only}

As noted in the previous subsection, if we keep track of all demes in which the additional ancestral lineage at a given locus resides at any given time in the past, then the MCG is a complicated object. To remedy this, we approximate the full marginal genealogy by just recording the time until absorption, as well as the deme in which the ancestral lineage resides at the time of absorption and also the absorbing haplotype. The reduced MCG at locus $\locusIdx$ is thus given by a triplet of random variables $(\absDemeRV{\locusIdx},\absTimeRV{\locusIdx},\absHapRV{\locusIdx}) \in \demeSpace \times \R_{\geq 0} \times \hapSpace$, that is the deme of absorption $\absDemeRV{\locusIdx}$, the absorption time $\absTimeRV{\locusIdx}$, and the absorbing haplotype $\absHapRV{\locusIdx}$. Henceforward, we use $\hStateSpace$ to denote $\demeSpace \times \R_{\geq 0} \times \hapSpace$. 

Now, observe that the marginal migration dynamics of the ancestral lineage at a single locus can be described by a continuous-time Markov chain with a finite state space. The states can be divided in two groups: one state for each deme denoting \emph{residence} in that deme before being absorbed into the trunk, and another one for each deme to represent being \emph{absorbed} into a lineage of the trunk in the given deme at some previous time. We denote the set of states by $\{1,\ldots,\numDemes,\absState{1},\ldots,\absState{\numDemes}\}$, where, for $1\leq i\leq g$, state $i$ denotes residence in deme $i$, and state $\absState{i}$ denotes absorption in deme $i$. The dynamics of the Markov chain is given by the $(2\numDemes\times 2\numDemes)$-dimensional block-specified rate matrix
\begin{equation}
	\nMatrix = \left( \begin{matrix}
		\migMatrix - \absMatrix	& \absMatrix\\
		\mathbf{0} 	& \mathbf{0}
	\end{matrix} \right),
\end{equation}
where $\mathbf{0}$ is a $(\numDemes\times\numDemes)$-dimensional matrix of zeros, $\migMatrix$ is the $(\numDemes\times\numDemes)$-dimensional matrix of migration rates which govern the transitions between the first group of states (the non-absorbed states), and $\absMatrix$ is the $(\numDemes\times\numDemes)$-dimensional diagonal matrix
\begin{equation}
	\absMatrix = \left( \begin{matrix}
		\relDemeSize{1}^{-1} \nSizeDeme{1}	& \cdots	& 0 \\
		\vdots	& \ddots	& \vdots \\
		0	& \cdots	& \relDemeSize{\numDemes}^{-1}\nSizeDeme{\numDemes}
	\end{matrix} \right)
\end{equation}
which governs the transition into the second group (the absorbed states). The diagonal form ensures that the absorbed state $\absState{i}$ can be reached only if the ancestral lineage currently resides in deme $i$.  Also, note that absorption is proportional to the inverse of relative size $\relDemeSize{i}^{-1}$ of deme $i$, as well as the number of trunk-lineages $\nSizeDeme{i}$ in deme $i$. Because the absorbing states are also absorbing in the context of the Markov chain, the rows of $\nMatrix$ corresponding to these states are set to zero.

The process generating the conditional genealogy for the whole additional haplotype proceeds sequentially along the haplotype, and thus admits a natural interpretation in an HMM framework, where the MCG at a given locus is the hidden state and the allele of the additionally sampled haplotype at this locus is the emitted symbol. We now describe the initial density, the transition density, and the emission probability.

\subsubsection{Initial density}
\label{sec_init_density}
Standard theory of Markov chains yields that the probabilities of interest for the initial density can be found in the respective entries of the transition semigroup. If the additional haplotype is sampled at present in deme $\alpha$, the probability of residing in deme $i$ and not being absorbed more recently than time $t$ into the past is $(e^{ \absTime{} \nMatrix})_{\addDeme,i}$. On the other hand, the cumulative probability of being absorbed in deme $i$ more recently than time $t$ is given by $(e^{ \absTime{} \nMatrix })_{\addDeme,\absState{i}}$. Thus, the initial density of state $\hState{} = (\absDeme{}, \absHap{}, \absTime{})$, that is the density of being absorbed in deme $\absDeme{}$ at time $\absTime{}$ into the trunk-lineage of haplotype $\absHap{}$, is given by the derivative of the latter matrix exponential:
\begin{equation}\label{eq_def_initial_density}
	\begin{split}
		\iDensity^{(\nCfg)} (\hState{}) & = \frac{d}{d\absTime{}} \P\{\absDemeRV{} = \absDeme{}, \absTimeRV{} \leq \absTime{}, \absHapRV{} = \absHap{} \}\\
			& = \frac{d}{d\absTime{}} \bigg\{ \frac{\nCmp{\absDeme{}}{\absHap{}}}{\nSizeDeme{\absDeme{}}} \big(e^{\absTime{} \nMatrix}\big)_{\alpha,\absState{\absDeme{}}} \bigg\}\\
			& = \frac{\nCmp{\absDeme{}}{\absHap{}}}{\nSizeDeme{\absDeme{}}} \big(\nMatrix e^{\absTime{} \nMatrix}\big)_{\alpha,\absState{\absDeme{}}}.\\
	\end{split}
\end{equation}
The factor $\nCmp{\absDeme{}}{\absHap{}}/\nSizeDeme{\absDeme{}}$ comes from the fact that absorption into a specific lineage of the trunk is uniform amongst those present in deme $\absDeme{}$.

\subsubsection{Transition density}
The density for transition from locus $\locusIdx-1$ to $\locusIdx$ using the full MCG, described in Section~\ref{sec_smc}, conditions on the full migration history of the lineage of the additionally sampled haplotype at locus $\locusIdx-1$. Thus, at the time of a possible recombination event, all demes up to this event, including the deme where the event takes place are determined. If only the time and the deme of absorption are recorded, then the deme in which the ancestral lineage resides at the time of the recombination event is a random variable with a distribution determined by the dynamics of the Markov chain. Let $\breakDemeRV{\breakTime}$ denote the random deme in which the ancestral lineage of the additional haplotype resides at the time $\breakTime$ of the recombination event.  Then, for $\breakDeme \in \demeSpace$, the distribution is given by
\begin{equation}\label{eq_conditional_deme_distribution}
		\P \{ \breakDemeRV{\breakTime} = \breakDeme \mid \absTimeRV{\locusIdx-1} = \absTime{\locusIdx-1}, \absDemeRV{\locusIdx-1} = \absDeme{\locusIdx-1}\} = \frac{\big[ e^{ \breakTime \nMatrix}\big]_{\addDeme,\breakDeme} \big[ \nMatrix e^{ (\absTime{\locusIdx-1} - \breakTime) \nMatrix}\big]_{\breakDeme,\absState{\absDeme{\locusIdx-1}}}}{\big[ \nMatrix e^{\absTime{\locusIdx-1} \nMatrix}\big]_{\addDeme,\absState{\absDeme{\locusIdx-1}}}}.
\end{equation}
The transition from $\hState{\locusIdx-1}$ to $\hState{\locusIdx}$ is now given as follows:  The time $\breakTime$ of the potential recombination event is chosen according to an exponential distribution with rate $\recRate{(\locusIdx-1,\locusIdx)}$.  
If $\breakTime \geq \absTime{\locusIdx-1}$, no recombination occurs and the MCG $\hState{\locusIdx}$ is identical to $\hState{\locusIdx-1}$. If $\breakTime < \absTime{\locusIdx-1}$, then recombination occurs and we use \eqref{eq_conditional_deme_distribution} to determine the probability that the ancestral lineage resided in a certain deme at the time of the recombination event. The ancestral lineage at locus $\locusIdx$ proceeds from time $\breakTime$ in this deme and is again subject to migration according to the dynamics of the Markov chain governed by rate matrix $\nMatrix$, until it is absorbed. Integrating over the possible times of the recombination event and summing over the different possible demes yields the following transition density of the hidden state $\hState{\locusIdx}$ at locus $\locusIdx$, given $\hState{\locusIdx-1}$ at the previous locus:
\begin{equation}\label{eq_def_transition_density}
	\begin{split}
		\transDensity^{(\nCfg)}_{\recRate{\breakPoint}}(\hState{\locusIdx} \vert \hState{\locusIdx-1}) & = \P\{ \absDemeRV{\locusIdx} = \absDeme{\locusIdx}, \absTimeRV{\locusIdx} \in d\absTime{\locusIdx}, \absHapRV{\locusIdx} = \absHap{\locusIdx} \vert \absDemeRV{\locusIdx-1} = \absDeme{\locusIdx-1}, \absTimeRV{\locusIdx-1} = \absTime{\locusIdx-1}, \absHapRV{\locusIdx-1} = \absHap{\locusIdx-1}\}\\
			& =  e^{-\recRate{\breakPoint} \absTime{\locusIdx-1}} \delta_{\hState{\locusIdx}, \hState{\locusIdx-1}}\\
			& \qquad + \frac{\nCmp{\absDeme{\locusIdx}}{\absHap{\locusIdx}}}{\nSizeDeme{\absDeme{\locusIdx}}} \int_{\breakTime = 0}^{\absTime{\locusIdx-1} \wedge \absTime{\locusIdx}} e^{-\recRate{\breakPoint} \breakTime} \sum_{\breakDeme \in \demeSpace} \frac{\big[ e^{ \breakTime \nMatrix}\big]_{\addDeme,\breakDeme} \big[ \nMatrix e^{ (\absTime{\locusIdx-1} - \breakTime) \nMatrix}\big]_{\breakDeme,\absState{\absDeme{\locusIdx-1}}}}{\big[ \nMatrix e^{\absTime{\locusIdx-1} \nMatrix}\big]_{\addDeme,\absState{\absDeme{\locusIdx-1}}}} \big[ \nMatrix e^{ (\absTime{\locusIdx} - \breakTime) \nMatrix}\big]_{\breakDeme,\absState{\absDeme{\locusIdx}}} d\breakTime,\\
	\end{split}
\end{equation}
where $\absTime{\locusIdx-1} \wedge \absTime{\locusIdx}$ denotes the minimum of $\absTime{\locusIdx-1}$ and $\absTime{\locusIdx}$.

\subsubsection{Emission probability}
\label{sec_emission_prob}

Since the mutation rate does not depend on the deme in which the ancestral lineage of the additional haplotype resides, the emission probability at locus $\locusIdx$ only depends on the absorption time $\absTime{\locusIdx}$ and the allele of the absorbing haplotype at that locus $\absHap{\locusIdx}[\locusIdx]$. As described above, a Poisson number (with mean $\absTime{\locusIdx}\mutRate{\locusIdx}$) of mutation events is placed onto the MCG $\hState{\locusIdx}$ and the ``ancestral'' allele is propagated to the present according to the mutation transition matrix $\mutMatrix{\locusIdx}$.
Thus, the probability that the allele of the additional haplotype is $\addHap[\locusIdx]$, given the hidden state $\hState{\locusIdx}$, can be written as the following matrix exponential:
\begin{equation}\label{eq_def_emission_probability}
	\begin{split}
		\emDensity^{(\nCfg)}_{\mutRate{\locusIdx}} (\addHap[\locusIdx] \mid \hState{\locusIdx}) & = \P \{ \addHapRV_\locusIdx = \addHap[\locusIdx] \mid \absDemeRV{\locusIdx} = \absDeme{\locusIdx}, \absTimeRV{\locusIdx} = \absTime{\locusIdx}, \absHapRV{\locusIdx} = \absHap{\locusIdx} \}\\
		 	& = \big[ e^{\absTime{\locusIdx} \mutRate{\locusIdx} (\mutMatrix{\locusIdx} - \1)}\big]_{\absHap{\locusIdx}[\locusIdx],\addHap[\locusIdx]},
	\end{split}
\end{equation}
where $\addHapRV_\locusIdx$ denotes the random allele emitted at locus $\locusIdx$.

\subsubsection{A hidden Markov model formulation to compute the CSP}

Using the quantities introduced in the previous Sections, we can now employ the forward algorithm for this HMM with continuous hidden state space \citep{Cappe2005} by defining the quantity $\fwDensAbsOnly^\addHap_\locusIdx (\cdot)$ recursively. The base case for the first locus is
\begin{equation}
	\fwDensAbsOnly^\addHap_1 (\hState{1}) = \emDensity^{(\nCfg)}_{\mutRate{1}} (\addHap[1] \vert \hState{1}) \cdot \iDensity^{(\nCfg)} (\hState{1}),
\end{equation}
and the recursive step for the transition from locus $\locusIdx-1$ to $\locusIdx$ is
\begin{equation}
	\begin{split}
		\fwDensAbsOnly^\addHap_\locusIdx (\hState{\locusIdx}) = \emDensity^{(\nCfg)}_{\mutRate{\locusIdx}} (\addHap[\locusIdx] \vert \hState{\locusIdx}) \cdot \int \transDensity^{(\nCfg)}_{\recRate{\breakPoint}}(\hState{\locusIdx} \vert \hState{\locusIdx-1}) \fwDensAbsOnly^\addHap_{\locusIdx-1} (\hState{\locusIdx-1}) d\hState{\locusIdx-1},
	\end{split}
\end{equation}
where $\breakPoint = (\locusIdx-1,\locusIdx)$. Finally, the CSP $\pihatmigsmcAO (\addHap \vert {\nCfg} )$ of an additional haplotype $\addHap$ given the already observed sample configuration $\nCfg$ is given by:
\begin{equation}
	\pihatmigsmcAO (\addHap \vert {\nCfg} ) = \int\fwDensAbsOnly^\addHap_\numLoci (\hState{\numLoci}) d\hState{\numLoci}.
\end{equation}
The sequentially Markov genealogical process corresponding to the CSD $\pihatmigsmcAO$ is illustrated in Figure~\ref{fig_csd_absorption_only}.

Note that the dynamics of the Markov chain on the hidden states is reversible with respect to the initial density, i.e., 
\begin{equation}
	\transDensity^{(\nCfg)}_{\recRate{}}(\hState{}' \vert \hState{}) \iDensity^{(\nCfg)} (\hState{}) = \transDensity^{(\nCfg)}_{\recRate{}}(\hState{} \vert \hState{}') \iDensity^{(\nCfg)} (\hState{}')
\end{equation}
holds for all recombination rates $\recRate{} \in \R_{\geq 0}$ and hidden states $\hState{}, \hState{}' \in \hStateSpace$. Thus, the initial density $\iDensity^{(\nCfg)} (\cdot)$ is in fact the stationary distribution of the Markov chain on the hidden state space. Reversibility also ensures that the CSP computed starting at the first locus and proceeding forward is the same as the CSP computed when starting at the final locus and proceeding backward.

\subsection{Discretizing time}

The reduced hidden state space of the HMM introduced in the previous subsection yields an approximation to the full sequentially Markov conditional sampling distribution. However, the hidden state space (in particular the absorption time) is continuous, making implementation with standard (discrete) HMM methodology impossible. Thus, as in \citet{Paul2011}, we propose a further approximation, by discretizing the positive real line into a finite number of intervals and recording the interval that the absorption time falls into. Formally, this corresponds to the approximation that the transition density and emission probability are equal for times that belong to the same interval.

To this end, assume that $0 = \partPoint{0} < \partPoint{1} < \cdots < \partPoint{\numPart} = \infty$ is a finite, strictly increasing sequence in $\R_{\geq 0} \cup \{\infty\}$ such that $\part = \{\partInt{j} = [\partPoint{j-1},\partPoint{j})\}_{j=1,\ldots,\numPart}$ is a partition of $\R_{\geq 0}$ into $\numPart$ intervals. We denote the discretized hidden state space $\demeSpace \times \{1,\ldots,\numPart \} \times \hapSpace$ by $\hStateSpaceDisc$ and the hidden states by $\hStateDisc{} = (\absDeme{}, \partIdx{}, \absHap{}) \in \hStateSpaceDisc$, where $\partIdx{}$ is the index of the interval of absorption. As before, $\absDeme{}$ denotes the deme and $\absHap{}$ the trunk lineage of absorption. Based on the partition $\part$, denote the discretized version of the initial density as
\begin{equation}\label{eq_def_disc_initial_probability}
	\begin{split}
		\iDensityDisc^{(\nCfg)} (\hStateDisc{}) & = \P\{\absDemeRV{} = \absDeme{}, \absTimeRV{} \in \partInt{\partIdx{}}, \absHapRV{} = \absHap{} \}\\
			& = \int_{\partInt{\partIdx{}}} \iDensity^{(\nCfg)} (\absDeme{},\absTime{},\absHap{}) d\absTime{}\\
			& = \margProb{\absDeme{}}{\partIdx{}} \cdot  \frac{\nCmp{\absDeme{}}{\absHap{}}}{\nSizeDeme{\absDeme{}}},
	\end{split}
\end{equation}
where 
\begin{equation}\label{eq_marginal_prob}
	\margProb{\absDeme{}}{\partIdx{}} = \int_{\absTime{} \in \partInt{\partIdx{}}} \big(\nMatrix e^{\absTime{} \nMatrix}\big)_{\alpha,\absState{\absDeme{}}} d\absTime{} = \big(e^{ \partPoint{\partIdx{}} \nMatrix}\big)_{\alpha,\absState{\absDeme{}}}  - \big(e^{ \partPoint{\partIdx{}-1} \nMatrix}\big)_{\alpha,\absState{\absDeme{}}}.
\end{equation}
Note that the event $\{\absTimeRV{} \in \partInt{\partIdx{}}\}$ encodes that we only record the time interval in which absorption happens.

Similarly, we can derive the discretized version of the transition density as
\begin{equation}\label{eq_def_disc_transition_probability}
	\begin{split}
		\transDensityDisc^{(\nCfg)}_{\recRate{\breakPoint}}(\hStateDisc{\locusIdx} \vert \hStateDisc{\locusIdx-1}) & = \P\{ \absDemeRV{\locusIdx} = \absDeme{\locusIdx}, \absTimeRV{\locusIdx} \in \partInt{\partIdx{\locusIdx}}, \absHapRV{\locusIdx} = \absHap{\locusIdx} \vert \absDemeRV{\locusIdx-1} = \absDeme{\locusIdx-1}, \absTimeRV{\locusIdx-1} \in \partInt{\partIdx{\locusIdx-1}}, \absHapRV{\locusIdx-1} = \absHap{\locusIdx-1}\}\\
			& = \frac{1}{\iDensityDisc^{(\nCfg)} (\hStateDisc{\locusIdx-1})} \int_{\partInt{\partIdx{\locusIdx}}} \int_{\partInt{\partIdx{\locusIdx-1}}}  \transDensity^{(\nCfg)}_{\recRate{\breakPoint}}(\absDeme{\locusIdx},\absTime{\locusIdx},\absHap{\locusIdx} \vert \absDeme{\locusIdx-1},\absTime{\locusIdx-1},\absHap{\locusIdx-1}) \iDensity^{(\nCfg)} (\absDeme{\locusIdx-1},\absTime{\locusIdx-1},\absHap{\locusIdx-1}) d\absTime{\locusIdx-1} d\absTime{\locusIdx}\\
			& = \noRecoTrans{\absDeme{\locusIdx-1}}{\partIdx{\locusIdx-1}}{\recRate{\breakPoint}} \cdot \delta_{\hState{\locusIdx-1}, \hState{\locusIdx}} + \recoTrans{\absDeme{\locusIdx-1}}{\partIdx{\locusIdx-1}}{\absDeme{\locusIdx}}{\partIdx{\locusIdx}}{\recRate{\breakPoint}}  \cdot \frac{\nCmp{\absDeme{\locusIdx}}{\absHap{\locusIdx}}}{\nSizeDeme{\absDeme{\locusIdx}}},
	\end{split}
\end{equation} 
where explicit expressions of $\noRecoTrans{\absDeme{\locusIdx-1}}{\partIdx{\locusIdx-1}}{\recRate{\breakPoint}}$ and $\recoTrans{\absDeme{\locusIdx-1}}{\partIdx{\locusIdx-1}}{\absDeme{\locusIdx}}{\partIdx{\locusIdx}}{\recRate{\breakPoint}}$ are shown in Appendix~\ref{app_discretized_probabilities}.  Note that we again only record the intervals containing the absorption times at locus $\locusIdx-1$ and $\locusIdx$. 

Finally, the emission probabilities in the discretized HMM can be obtained via
\begin{equation}\label{eq_def_disc_emission_probability}
	\begin{split}
		\emDensityDisc^{(\nCfg)}_{\mutRate{\locusIdx}} (\addHap[\locusIdx] \vert \hStateDisc{\locusIdx}) & = \P \{ \addHapRV_\locusIdx = \addHap[\locusIdx] \vert \absDemeRV{\locusIdx} = \absDeme{\locusIdx}, \absTimeRV{\locusIdx} \in \partInt{\partIdx{\locusIdx}}, \absHapRV{\locusIdx} = \absHap{\locusIdx} \}\\
			& = \frac{1}{\iDensityDisc^{(\nCfg)} (\hStateDisc{\locusIdx})} \int_{\partInt{\partIdx{\locusIdx}}} \emDensity^{(\nCfg)}_{\mutRate{\locusIdx}} (\addHap[\locusIdx] \vert \absDeme{\locusIdx},\absTime{\locusIdx},\absHap{\locusIdx}) \iDensity^{(\nCfg)} (\absDeme{\locusIdx},\absTime{\locusIdx},\absHap{\locusIdx}) d\absTime{\locusIdx},\\
	\end{split}
\end{equation}
and we again provide a more explicit form of this quantity in Appendix~\ref{app_discretized_probabilities}. Note that the emission probability~\eqref{eq_def_emission_probability} in the continuous case is only dependent on the time of absorption and the allele that the absorbing haplotype bears at the given locus. The discretized analog~\eqref{eq_def_disc_emission_probability} on the other hand also depends on the deme that the absorbing haplotype resides in. This is due to the fact that the latter conditions on being absorbed at any point in a given time interval, and since the rate of absorption during that interval depends on the deme, this dependence enters expression~\eqref{eq_def_disc_emission_probability}.

With the state space discretized, the CSP can be computed via the standard forward algorithm for HMMs \citep{Cappe2005}. Thus, we define the quantity $\forwardVar^\addHap_\locusIdx (\hStateDisc{\locusIdx})$ recursively along loci. At the first locus, we have
\begin{equation}
	\forwardVar^\addHap_1 (\hStateDisc{1}) = \emDensityDisc^{(\nCfg)}_{\mutRate{1}} (\addHap[1] \vert \hStateDisc{1}) \cdot \iDensityDisc^{(\nCfg)} (\hStateDisc{1}).
\end{equation}
The transition from locus $\locusIdx-1$ to locus $\locusIdx$ is given by
\begin{equation}
	\forwardVar^\addHap_\locusIdx (\hStateDisc{\locusIdx}) = \emDensityDisc^{(\nCfg)}_{\mutRate{\locusIdx}} (\addHap[\locusIdx] \vert \hStateDisc{\locusIdx}) \cdot \sum_{\hStateDisc{\locusIdx-1} \in \hStateSpaceDisc} \transDensityDisc^{(\nCfg)}_{\recRate{\breakPoint}}(\hStateDisc{\locusIdx} \vert \hStateDisc{\locusIdx-1}) \forwardVar^\addHap_{\locusIdx-1} (\hStateDisc{\locusIdx-1}),
\end{equation}
and the probability of observing haplotype $\addHap$ under the discretized HMM is given by 
\begin{equation}
	\pihatmigsmcDisc(\addHap \vert {\nCfg}) := \sum_{\hStateDisc{\numLoci} \in \hStateSpaceDisc} \forwardVar^\addHap_{\numLoci} (\hStateDisc{\numLoci}),
\end{equation}
which provides an approximation to $\pihatmigsmcAO(\addHap \vert {\nCfg})$.

\begin{remark} \hspace{1mm}
	\begin{enumerate}
		\item In \citet{Paul2011}, the authors advocate using a discretization based on points obtained from Gaussian quadrature. However, we obtained numerically more stable results when using a logarithmic discretization, that is $\partPoint{i} = -(1/r) \log((\numPart-i)/i)$, where $r$ is the harmonic mean of the absorption rates in each deme.
		\item The runtime of the standard implementation of the forward algorithm for HMMs described in the previous paragraph is quadratic in the number of hidden states. In \citet{Paul2011}, the authors describe a straightforward implementation of their model that leads to a better bound on the runtime. Since our transition density is of similar form, a similar improvement can be applied here.
	\end{enumerate}
\end{remark}

\section{Application: Estimating migration rates} \label{sec_estimate_migration_rates}
To demonstrate the utility of our approximate CSD $\pihatmigsmcDisc$, we considered estimating migration rates for data simulated under the full coalescent with recombination and migration. In particular, we simulated data for $\numLoci = 10^4$ bi-allelic loci. For simplicity, we set $\mutRate{\loc} = 5\times10^{-2}$ and $\mutMatrix{\loc}=\left(\begin{smallmatrix}1/2 & 1/2\\ 1/2&1/2\end{smallmatrix}\right)$ for all $\loc\in\locSet$, and $\recRate{\breakPoint} = 5\times10^{-2}$ for all $\breakPoint\in\breakPointSet$. We considered a structured population with two demes (i.e., $\demeSpace = \{1,2\}$), and set $\relDemeSize{1} = \relDemeSize{2} = 0.5$ and $\migRate{1}{2} = \migRate{2}{1} = \migRateSingle$. For each value of $\migRateSingle\in\{0.01,0.10,1.00,10.0\}$, we generated $100$ datasets with $\nSizeDeme{1} = \nSizeDeme{2} = 10$ individuals in each of the two demes.

Observe that the per-individual mutation and recombination rates are thus both approximately $10^4 \cdot 5\times10^{-2} = 5 \times 10^2$. In humans, for which average per-base mutation and recombination rates are on the order of $10^{-3}$, these values correspond to a genomic sequence on the order of $500$~kb. We thus reason that the haplotypes we simulated are representative of a relatively longer genomic sequence that has been ``compressed'', for reasons of computational efficiency, into $10^4$ loci. Further, we chose the range of migration rates to be compliant with recent estimates in humans \citep{Gutenkunst2009,Gravel2011}, as well as Drosophila \citep{Wang2010}.

\begin{figure}[t]
	\centering
	\psfrag{ylab}[tc][Bc][1][0]{Scaled log-likelihood}
	\psfrag{xlab}[Bc][tc][1][0]{Migration rate, $\migRateSingle$}
	
	\psfrag{0.0}[Bc][cc][1][0]{$0.0$}
	\psfrag{0.1}[Bc][cc][1][0]{$0.1$}
	\psfrag{0.2}[Bc][cc][1][0]{$0.2$}
	\psfrag{0.3}[Bc][cc][1][0]{$0.3$}
	\psfrag{0.4}[Bc][cc][1][0]{$0.4$}
	
	\psfrag{pac}[Bl][Bl][1][0]{PAC}
	\psfrag{pair}[Bl][Bl][1][0]{PCL}
	\psfrag{spac}[Bl][Bl][1][0]{LCL}     
	
	\subfigure[] {		
		\includegraphics[angle=270,trim=20mm 15mm 5mm 5mm, clip,width=.45\textwidth]{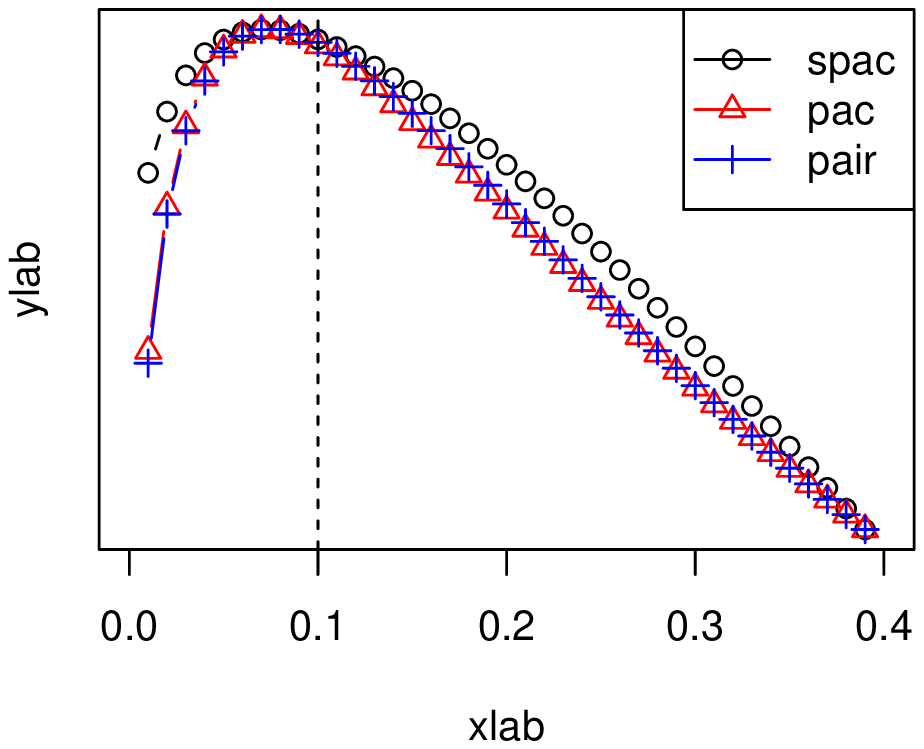}
		\label{fig:miglikgood}
	}
	\subfigure[] {		
		\includegraphics[angle=270,trim=20mm 15mm 5mm 5mm, clip,width=.45\textwidth]{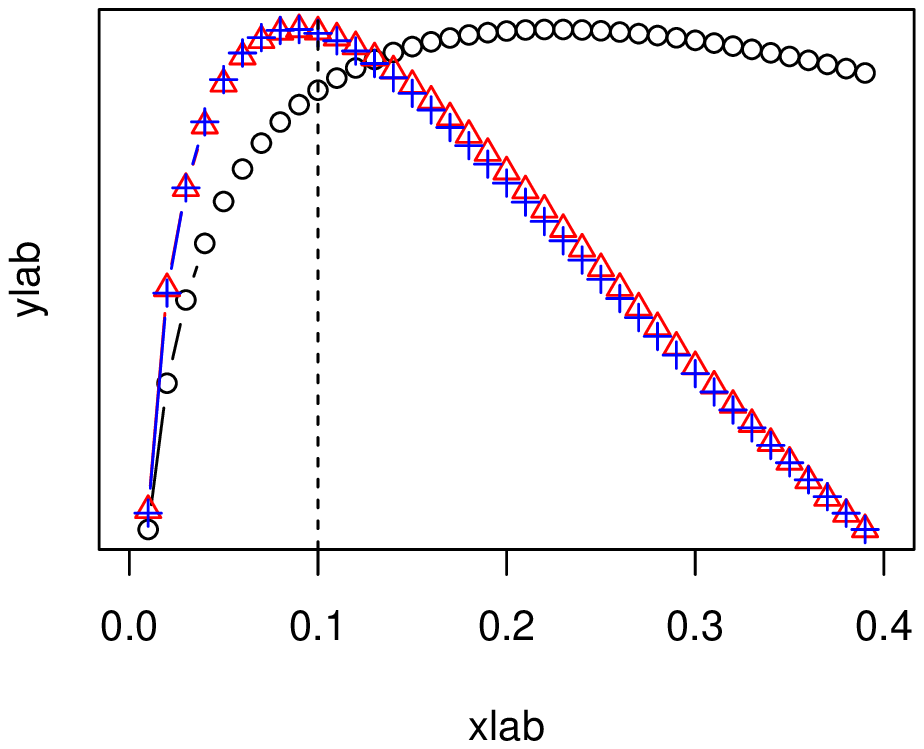}
		\label{fig:miglikbad}
	}
	\caption{\small Re-scaled log likelihood surfaces for two sample configurations (generated for $\migRateSingle = 0.10$, indicated by a vertical line in the plots), and for each of the three approximate likelihood formulations (LCL, PAC, PCL) described in the text. In both cases, the likelihoods are computed using the true values of $\mutRateSingle = 5\times10^{-2}$ and $\recRateSingle = 5\times10^{-2}$. \subref{fig:miglikgood} A case for which all of the likelihood surfaces are similar \subref{fig:miglikbad} A case for which the LCL likelihood surface is substantially different than the likelihood surfaces for PAC and PCL}
	\label{fig:miglik}
\end{figure}

We considered three approximate/composite likelihood formulations that make use of the CSD. Let $\nCfg$ be a particular sample configuration of $n$ haplotypes, and write $\nCfg = \sum_{i=1}^{\nSize} \sngConfig{\deme_i,\hap_i}$.
\begin{description}
	\item[LCL:] In the \emph{Leave-one-out Composite Likelihood}, the likelihood is approximated as a product of CSPs with each the result of removing a single haplotype from the sample configuration:
	\begin{equation*}
		\LCL(\nCfg) = \bigg[\prod_{i=1}^{\nSize} \pihatmigsmcDisc(\sngConfig{\deme_i,\hap_i} | \nCfg - \sngConfig{\deme_i,\hap_i})\bigg]^{1/\nSize}.
	\end{equation*}
	\item[PAC:] In the popular \emph{Product of Approximate Conditionals} framework \citep{Li2003}, the likelihood is approximated by a conditional decomposition, averaged over 20 random permutations $\{\sigma_j\}$ of the haplotypes (this number of permutations is as suggested by \citeauthor{Li2003}). Defining $\sigma_j(\sngConfig{\deme_{i},\hap_{i}}) = \sngConfig{\deme_{\sigma_j(i)},\hap_{\sigma_j(i)}}$:
	\begin{equation*}
		\PAC(\nCfg) = \frac{1}{20}\sum_{j=1}^{20} \prod_{i=1}^{\nSize} \pihatmigsmcDisc\Big(\sigma_j(\sngConfig{\deme_{i},\hap_{i}}) | \nCfg - \sum_{i'=1}^i \sigma_j(\sngConfig{\deme_{i'},\hap_{i'}})\Big).
	\end{equation*}
	\item[PCL:] In the \emph{Pairwise Composite Likelihood}, the likelihood is approximated as a product of CSPs with each a single haplotype conditioned upon a single haplotype:
	\begin{equation*}
		\PCL(\nCfg) = \prod_{i=1}^{\nSize}\prod_{i'\neq i} \pihatmigsmcDisc(\sngConfig{\deme_i,\hap_i} | \sngConfig{\deme_{i'},\hap_{i'}}).
	\end{equation*}
\end{description}
We set the values of $\mutRateSingle$ and $\recRateSingle$ to the (true) values used for simulation, and considered the approximate likelihood surfaces for the parameter $\migRateSingle$. \fref{miglik} shows the surfaces for two example configurations (generated as described above) for $\migRateSingle = 0.10$. Perhaps most importantly, the likelihood surfaces appear to be unimodal and otherwise well-behaved. In \fref{miglikgood}, the likelihood curves are quite similar to one another, and the maximum likelihood occurs near the true parameter. This is not generally true, however, as evidenced by \fref{miglikbad}, for which the likelihood surface for the LCL method is substantially different than that of PAC and PCL. 

We also considered the behavior of the maximum likelihood estimate (MLE) under each of the likelihood approximations. For each simulated dataset, we computed, using golden section search, the MLE migration rate $\hat{\migRateSingle}$, and computed $\log_2(\hat{\migRateSingle}/\migRateSingle)$, where $\migRateSingle$ is the (true) migration rate used to generate the dataset. In this way, results for different values of $\migRateSingle$ are directly comparable; a correct estimate of the migration rate produces a value of $0$, and under- and overestimation by a factor of two produce values of $-1$ and $1$, respectively. To assess the performance of the MLEs based on the CSD developed in this paper, we also compared with estimates obtained from the widely-used test statistic $\fst$:
\begin{description}
	\item[$\fst$:] It can be shown that the migration rate in a symmetric island model with two sub-populations can be estimated by
	\begin{equation}
		\hat{\migRateSingle}(\nCfg) = \frac{1}{4}\bigg(\frac{1}{\fst(\nCfg)} - 1\bigg),
	\end{equation}
	where $\fst(\nCfg) = 1 - \piS(\nCfg) / \piT(\nCfg)$, with $\piS(\nCfg)$ denoting the average within-population diversity and $\piT(\nCfg)$ the overall diversity; c.f., \citet[Equation~(4)]{Charlesworth1998}. Note that, although \citeauthor{Charlesworth1998} discusses three different estimators for $\fst$, the corresponding migration rate estimators coincide in models where the sub-populations have equal weights.
\end{description}

\begin{figure}[t]
	\centering
	\psfrag{ylab}[tc][Bc][1][0]{$\log_2(\hat{m}/m)$}
	
	\psfrag{-3}[tc][cc][1][0]{$-3$}
	\psfrag{-2}[tc][cc][1][0]{$-2$}
	\psfrag{-1}[tc][cc][1][0]{$-1$}
	\psfrag{0}[tc][cc][1][0]{$-0$}
	\psfrag{1}[tc][cc][1][0]{$1$}
	\psfrag{2}[tc][cc][1][0]{$2$}
	\psfrag{3}[tc][cc][1][0]{$3$}
	
	\psfrag{pac}[Bc][cc][1][0]{PAC}
	\psfrag{pair}[Bc][cc][1][0]{PCL}
	\psfrag{spac}[Bc][cc][1][0]{LCL}
	
	\psfrag{ylab}[Bc][tc][1][0]{\footnotesize$\log_2(\hat{\migRateSingle}/\migRateSingle)$}
	
	\psfrag{-3}[cc][tc][1][0]{\footnotesize$-3$}
	\psfrag{-2}[cc][tc][1][0]{\footnotesize$-2$}
	\psfrag{-1}[cc][tc][1][0]{\footnotesize$-1$}
	\psfrag{0}[cc][tc][1][0]{\footnotesize$-0$}
	\psfrag{1}[cc][tc][1][0]{\footnotesize$1$}
	\psfrag{2}[cc][tc][1][0]{\footnotesize$2$}
	\psfrag{3}[cc][tc][1][0]{\footnotesize$3$}
	
	\psfrag{pac}[cc][Bc][1][0]{\footnotesize PAC}
	\psfrag{pair}[cc][Bc][1][0]{\footnotesize PCL}
	\psfrag{spac}[cc][Bc][1][0]{\footnotesize LCL}
	\psfrag{fst}[cc][Bc][1][0]{\footnotesize $\fst$}       

	\subfigure[] {		
		\includegraphics[width=.47\textwidth]{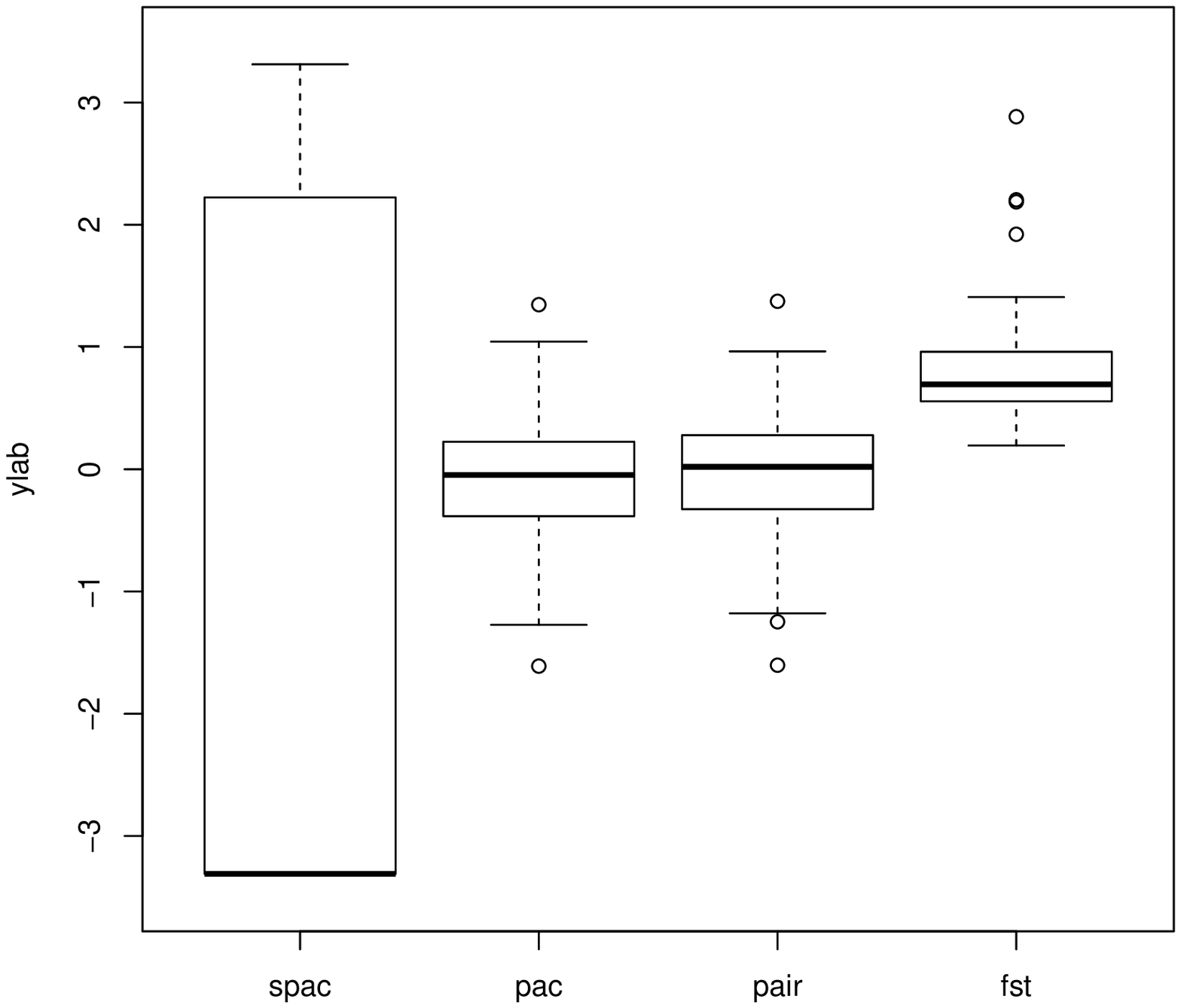}
		\label{fig:mig0-01}
	}
	\subfigure[] {		
		\includegraphics[width=.47\textwidth]{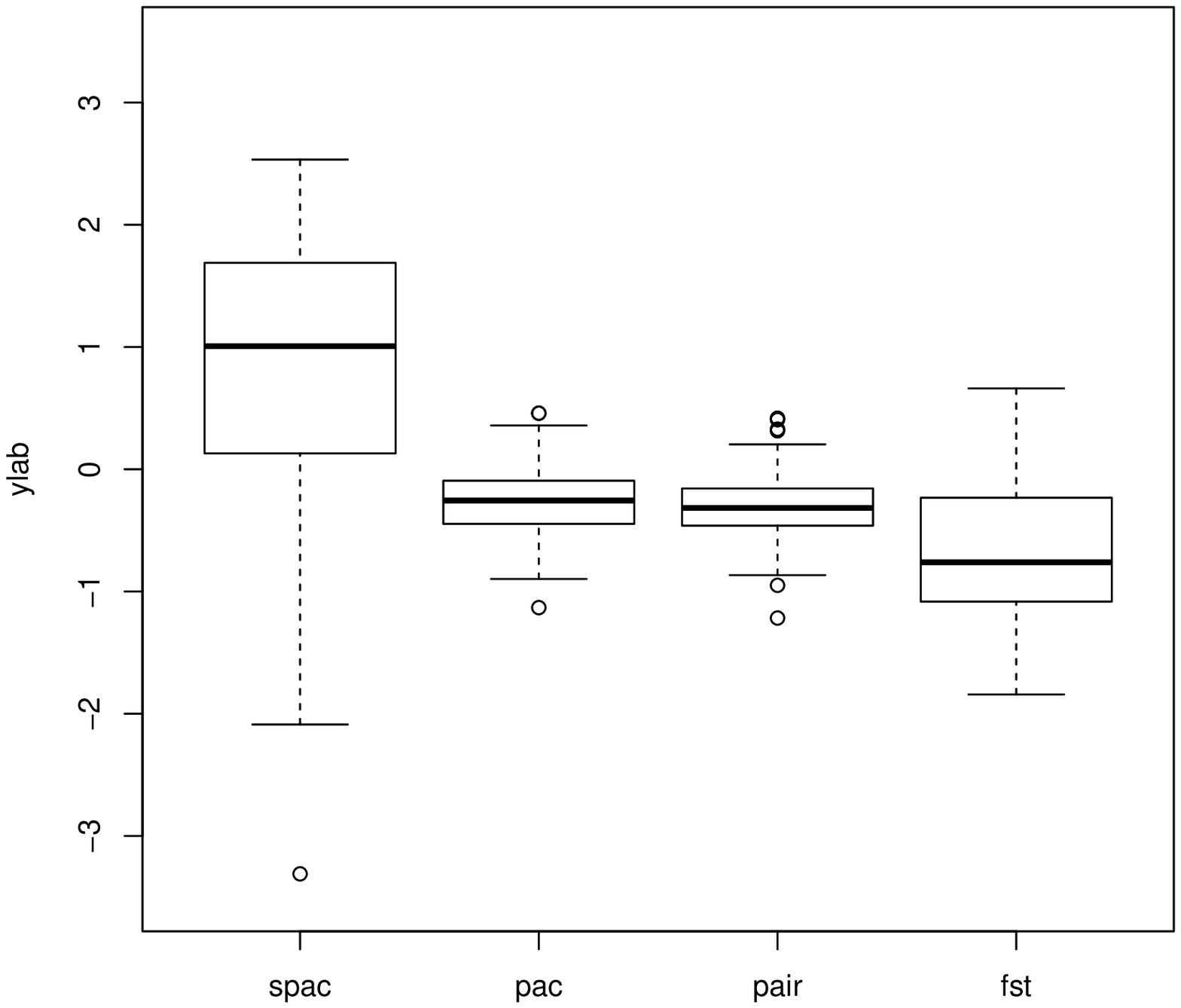}
		\label{fig:mig0-10}
	}
	\\
	\subfigure[] {		
		\includegraphics[width=.47\textwidth]{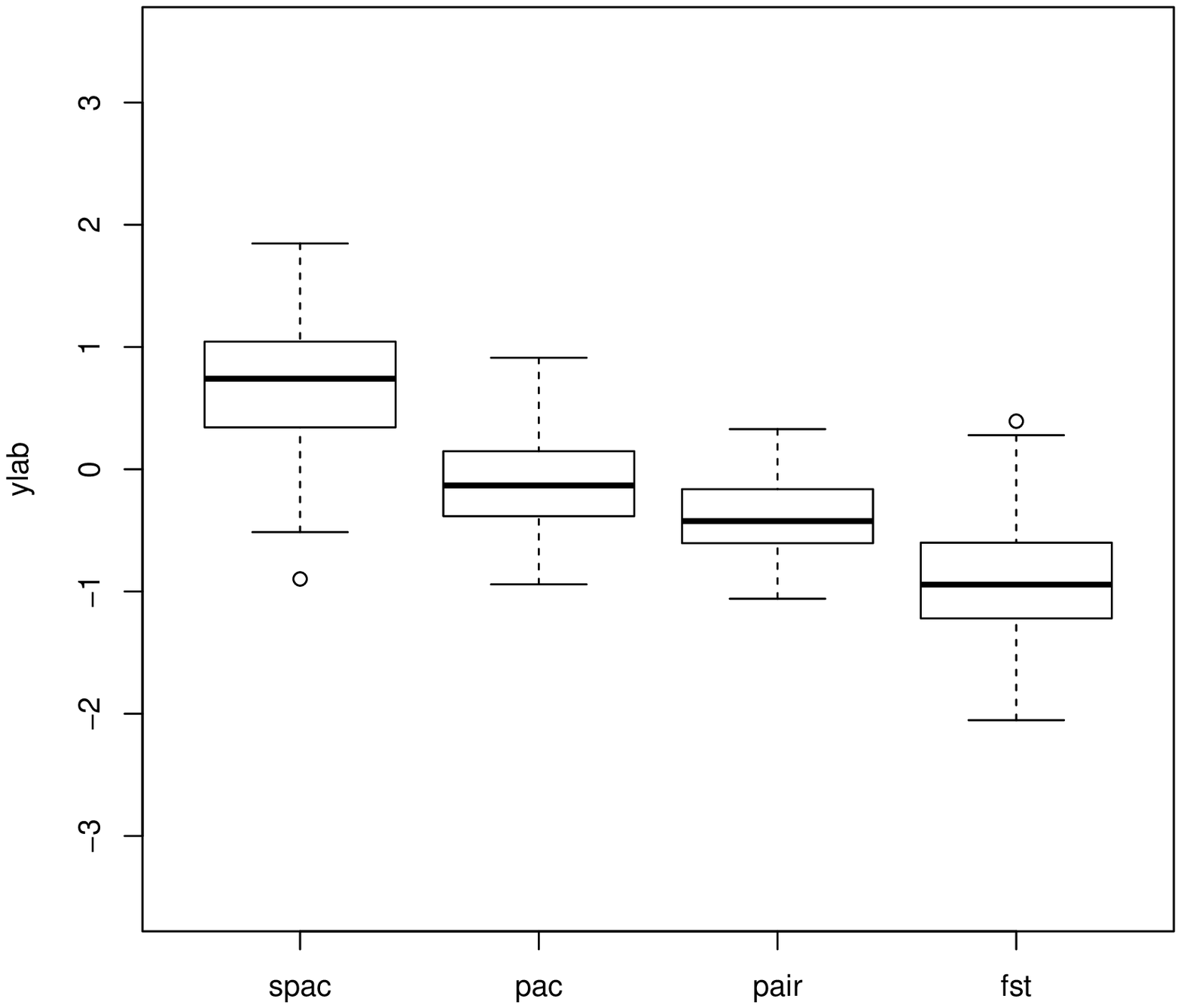}
		\label{fig:mig1-00}
	}
	\subfigure[] {
		\includegraphics[width=.47\textwidth]{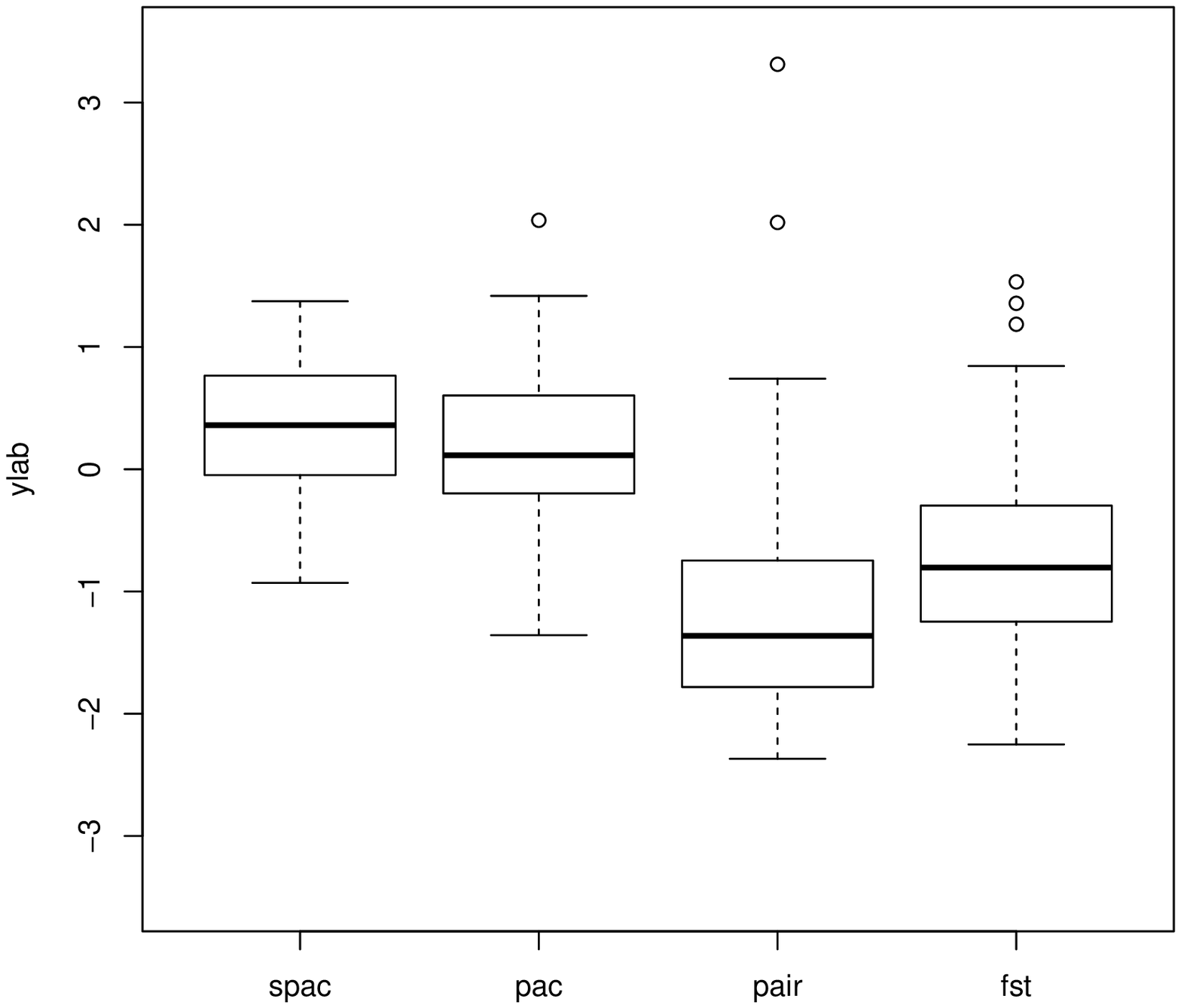}
		\label{fig:mig10-0}
	}
	\caption{\small Box plots (produced using the software package R, and including outliers) for the quantity $\log_2(\hat{\migRateSingle}/\migRateSingle)$ over $100$ samples, where $\hat{\migRateSingle}$ is the MLE under each of the three approximate likelihood formulations (LCL, PAC, PCL) or the $\fst$-based estimate as described in the text. The MLE values $\hat{\migRateSingle}$ were found using golden section search within the interval $(\migRateSingle \cdot 10^{-1}, \migRateSingle \cdot 10)$ \subref{fig:mig0-01} $\migRateSingle = 0.01$ \subref{fig:mig0-10} $\migRateSingle = 0.10$ \subref{fig:mig1-00} $\migRateSingle = 1.00$ \subref{fig:mig10-0} $\migRateSingle = 10.0$. Note that the median of the LCL estimator in~\subref{fig:mig0-01} lies on the lower bound of the interval, thus at least half of the estimates reach this bound and are most likely even smaller. }
	\label{fig:migest}
\end{figure}

For each true migration rate $\migRateSingle \in \{0.01,0.10,1.00,10.0\}$, box plots for the transformed MLE under each likelihood approximation and the $\fst$-based estimator are presented in \fref{migest}. Observe that the LCL-based MLE performs very poorly for $\migRateSingle = 0.01$ (see \fref{mig0-01}), consistently underestimating the true value; this may be because the final haplotype to be sampled is generally very similar to previously-sampled haplotypes within the deme, obviating the need for migration events within the conditional genealogy.  Intuitively, this effect should be diminished when the data are produced using larger migration rates, which does appears to be the case (see Figures~\ref{fig:mig0-10}, \ref{fig:mig1-00}, and \ref{fig:mig10-0}). 

On the other hand, the PCL-based MLE performed poorly for $\migRateSingle = 10.0$, again consistently underestimating the true value. This may be because, for large migration rates, there simply is not enough information in a pairwise analysis of the haplotypes to determine the true rate; intuitively, this effect should be diminished when the data are produced using smaller migration rates, relative to the rate of recombination. This is indeed the case, and in fact, for smaller migration rates, the PCL-based MLE is well-correlated with the PAC-based MLE (data not shown).

The PAC-based MLE appears not to suffer at either of these extremes. We speculate that this is because PAC incorporates both pairwise and higher-order terms, making it less susceptible to the problems we observe with the LCL- and PCL-based MLEs. We note that \citet{Li2003} came to a similar conclusion.

For low migration rates, the method based on $\fst$ consistently overestimates the true rates (see Figure~\ref{fig:mig0-01}), but shows a small variance.   For intermediate migration rates, on the other hand, it produces underestimates  (see Figures~\ref{fig:mig0-10} and~\ref{fig:mig1-00}), and the variance is larger than that of the PAC-based MLE.   The estimates for large migration rates (see Figure~\ref{fig:mig10-0}) are similarly biased, although the variance is comparable to the MLE methods in this case. Overall, the PAC-based estimation is quite accurate, demonstrating that, using the CSD $\pihatmigsmcDisc$, it is possible to attain accurate estimates of the migration rate.

\section{Discussion}
\label{sec_discussion}
Numerous applications in population genomics make use of the conditional sampling distribution, so developing accurate, efficiently computable CSDs for various population genetic models is of much interest.  Recently, we proposed an accurate sequentially Markov CSD that follows from approximating the diffusion process dual to the coalescent with recombination for a single panmictic population.  In this paper, we have extended that approach to incorporate subdivided population structure with migration, providing a novel CSD that facilitates computation and also admits a useful genealogical interpretation closely related to the structured coalescent with migration and recombination.  We believe that this extension will have several interesting applications, some of which we list below.

Recalling the applications of CSDs described in the introduction, we note that it is straightforward to apply $\pihatmigsmcDisc$ to annotate segments of distinct ancestry in individuals. As in \citet{Price2009}, the already observed configuration consists of the donor individuals from different populations. For low migration rates, the model underlying $\pihatmigsmcDisc$ leads naturally to the fact that, following a recombination event, the ancestral lineage at the next locus is more likely to get absorbed in the same deme, rather than switching demes by a migration event and then getting absorbed in a different deme. Whereas the method developed by \citet{Price2009} is applicable for recently admixed individuals, we expect our model to be more accurate in situations where the mixing of the populations happened over a long time through the continuous exchange of migrants.

Recall that \citet{Wegmann2011} estimated relative recombination rate variation in different populations based on ancestry switch-points in the chromosome detected using the method of \citet{Price2009}. As detailed above, our model can be extended to detect ancestry switch-points in populations that mixed over long periods of time. In such situations we expect that the segments of different ancestry detected by our method can be used in a similar fashion as in \citet{Wegmann2011} to analyze recombination rate variation in different subpopulations, when no strong recent admixture is evident.

Recently, \citet{Li2011} performed a related analysis of human demography. They used the SMC in the special case of a sample consisting of only two sequences, and thus were able to obtain explicit transition functions along the sequence, as we did for our CSD \citep{Paul2011}. \citeauthor{Li2011} incorporated changes in the size of the population into their model, thus allowing them to use the two sequences of a diploid individual to infer population size histories of different human populations. They do not explicitly account for population structure and migration in their analysis, but we believe that the methods developed in this paper could be readily incorporated into their model. In a similar study, \cite{Mailund2011} used a pair of sequences sampled from different populations in the SMC framework to estimate ancestral population sizes and splitting times in an isolation model. Again, we think it is possible to incorporate migration into the model using the ideas we developed in this paper.

\citet{Paul2012} have recently developed a framework to substantially increase the speed of computations involved in dealing with HMMs for next-generation sequence data, and they demonstrate their improvements in the model introduced by \citet{Paul2011}.  Utilizing the fact that whole genomic sequence data consists of long stretches without sequence variation in between SNPs, and that the observed variation can be described by a small number of haplotype blocks, they were able to decrease the computation time by several orders of magnitude. The same ideas can be applied to speedup our method, fostering the application of analyses like the one detailed in Section~\ref{sec_estimate_migration_rates} or similar applications to whole genome sequence data.

The CSDs developed in this paper assume that the structure underlying the population remains unchanged throughout the whole course of evolution. Furthermore the rates at which migrants are exchanged are assumed constant. This is mirrored by the fact that the Markov chain described by the matrix $\nMatrix$ is time homogeneous and the number of states does not change. For more realistic populations one would pose changes of the population structure and the population sizes at different points in the past, as well as varying rates of migration. The methods developed in this paper can be readily extended to scenarios where the structural parameters of the underlying demography are piecewise constant for given periods of time. This can be implemented by allowing the Markov chain, governing the absorption of the additional ancestral lineage, to be piecewise homogeneous. Except for the work of \citet{Davison2009} and \citet{Price2009}, we are not aware of any other CSDs that try to incorporate explicit population structure into the copying model. Such a CSD accounting for a more general demographic model would allow one to estimate more general demographic parameters like ancient population sizes and structure, as well as migration rates, and duration of periods of migration in certain isolation-with-migration scenarios, using, for example, the framework illustrated in Section~\ref{sec_estimate_migration_rates}, importance sampling \citep{Stephens2000,Fearnhead2001,Griffiths2008}, or other frameworks detailed in the introduction. \citet{Myers2008} show that demographic studies like \citet{Gutenkunst2009} and \citet{Gravel2011}, that rely exclusively on the frequency spectrum, can be limited in resolving demographic parameters, and methods, as the one developed in this paper, that explicitly incorporate linkage structure, might alleviate such problems.

\section*{Acknowledgments}
We thank John Kamm for many stimulating and fruitful discussions. This research is supported in part by a DFG Research Fellowship STE 2011/1-1 to M.S; an NIH National Research Service Award Trainee appointment on T32-HG00047 to JSP; and an NIH grant R01-GM094402, and a Packard Fellowship for Science and Engineering to Y.S.S.

\newpage
\appendix

\section{Diffusion approximation} \label{sec:app_diff_apprx}
We here provide a derivation of the sampling recursion using the diffusion generator approximation \citep{DeIorio2004a, DeIorio2004b, Griffiths2008, Paul2010}. The diffusion associated with the coalescent including migration and recombination has state given by the vector $\stateSpace = \big(\stateSpaceCmp{\deme}{\hap}\big)_{\deme\in\demeSpace, \hap\in\hapSpace}$ where $\stateSpaceCmp{\deme}{\hap}$ is the frequency of haplotype $\hap$ within deme $\deme$. The generator for the diffusion can then be written
\begin{equation*}
	\genTot f(\stateSpace) = \sum_{\deme\in\demeSpace \atop \hap\in\hapSpace} \genPart{\deme}{\hap} \frac{\partial}{\partial \stateSpaceCmp{\deme}{\hap}} f(\stateSpace),
\end{equation*}
where
\begin{align*}
	\genPart{\deme}{\hap} f(\stateSpace) 
	&= \frac{1}{2} \bigg\{ \stateSpaceCmp{\deme}{\hap} \sum_{\hap' \in \hapSpace} (\delta_{\hap,\hap'} - \stateSpaceCmp{\deme}{\hap'}) \relDemeSize{\deme}^{-1} \frac{\partial}{\partial \stateSpaceCmp{\deme}{\hap'}} f(\stateSpace) \\
	&\qquad + \sum_{\loc\in \locSet} \mutRate{\loc} \sum_{\allele \in \alleleSet{\loc}} \stateSpaceCmp{\deme}{\hapSub{\loc}{\allele}{\hap}} \big( \mutMatrix{\loc}_{\allele,\hap[\loc]} - \delta_{\hapSub{\loc}{\allele}{\hap},\hap}\big)  f(\stateSpace)\\
	&\qquad + \sum_{\breakPoint \in \breakPointSet} \recRate{\breakPoint} \Big( \sum_{\hap' \in \hapSpace}  \stateSpaceCmp{\deme}{\hapRec{\breakPoint}{\hap}{\hap'}} \stateSpaceCmp{\deme}{\hapRec{\breakPoint}{\hap'}{\hap}} - \stateSpaceCmp{\deme}{\hap}\Big) f(\stateSpace) \\
	&\qquad + \sum_{\deme' \neq \deme} \Big(\migRate{\deme}{\deme'} \stateSpaceCmp{\deme'}{\hap} - \migRateDeme{\deme} \stateSpaceCmp{\deme}{\hap}\Big) f(\stateSpace)  \bigg\},
\end{align*}
and $f$ is an arbitrary, bounded, twice-differentiable function with continuous second derivatives.

Denote the probability of obtaining (at stationarity) an ordered sample configuration $\nCfg$ by $\sampProb(\nCfg)$. Then $\sampProb(\nCfg) = \Ex[ \sampProb(\nCfg | \stateSpaceRV) ]$, where $\Ex$ denotes expectation with respect to the stationary distribution of the diffusion, $\stateSpaceRV$ denotes the random vector of frequencies, and $\sampProb(\nCfg | \stateSpace) = \prod_{\deme\in\demeSpace} \prod_{\hap\in\hapSpace} \stateSpaceCmp{\deme}{\hap}^{\nCmp{\deme}{\hap}}$. Finally, given an additional haplotype configuration $\addHapCfg$, the true conditional sampling probability is, by definition, $\pi(\addHapCfg | \nCfg) = \sampProb(\addHapCfg + \nCfg) / \sampProb(\nCfg)$.

By general diffusion theory, $\Ex[\genTot f(\stateSpaceRV)] = 0$. The diffusion generator approximation assumes the existence of a distribution, with associated expectation $\ExAppr$, such that the previous condition holds \emph{component-wise}; that is for each $\deme \in \demeSpace$ and $\hap\in\hapSpace$,
\begin{equation*} 
	\ExAppr\Bigg[\genPart{\deme}{\hap} \frac{\partial} {\partial \stateSpaceRVCmp{\deme}{\hap}} f(\stateSpaceRV)\Bigg] = 0.
\end{equation*}
By analogy to the exact case, we assume that the sampling distribution is approximated by $\sampProbAppr(\nCfg) = \ExAppr[\sampProb(\nCfg | \stateSpaceRV)]$, and define the approximate CSD $\pihatmig(\addHapCfg | \nCfg) = \sampProbAppr(\nCfg + \addHapCfg) / \sampProbAppr(\nCfg)$. Using the component-wise approximation above,
\begin{equation*}
\sum_{\deme\in\demeSpace}\sum_{\hap\in\hapSpace}\frac{\addHapCmp{\deme}{\hap}}{\addHapCmp{\deme}{\hap} + \nCmp{\deme}{\hap}} \ExAppr\bigg[\genPart{\deme}{\hap} \frac{\partial} {\partial \stateSpaceRVCmp{\deme}{\hap}} \sampProb(\nCfg + \addHapCfg | \stateSpaceRV) \bigg] = 0.
\end{equation*}
Using the expressions for $\genPart{\deme}{\hap}$ and $\sampProb(\nCfg + \addHapCfg | \stateSpaceRV)$, together with the definition $\sampProbAppr(\nCfg) = \ExAppr[\sampProb(\nCfg | \stateSpaceRV)]$, we obtain, after some algebra analogous to~\citet[Appendix]{Paul2010},
\begin{align*}
	\sampProbAppr(\nCfg+\addHapCfg) 
	&= \frac {1} {\mathcal{N}} \sum_{\deme\in\demeSpace \atop \hap\in\hapSpace} \addHapCmp{\deme}{\hap} \bigg[ (\nCmp{\deme}{\hap}+\addHapCmp{\deme}{\hap} -1) \relDemeSize{\deme}^{-1} \sampProbAppr(\nCfg + \addHapCfg - \sngConfig{\deme,\hap}) \\
	&\qquad+ \sum_{\loc\in\locSet} \mutRate{\loc} \sum_{\allele \in \alleleSet{\loc}} \mutMatrix{\loc}_{\allele,\hap[\loc]} \sampProbAppr(\nCfg + \addHapCfg - \sngConfig{\gamma,h} + \sngConfig{\gamma,\hapSub{\loc}{\allele}{\hap}}) \\ 
	&\qquad+ \sum_{\breakPoint \in \breakPointSet} \recRate{\breakPoint} \sum_{\hap' \in \hapSpace} \sampProbAppr(\nCfg + \addHapCfg - \sngConfig{\deme,\hap} + \sngConfig{\deme,\hapRec{\breakPoint}{\hap}{\hap'}} + \sngConfig{\deme,\hapRec{\breakPoint}{\hap'}{\hap}) } )\\
	&\qquad+ \sum_{\deme' \neq \deme} \migRate{\deme}{\deme'} \sampProbAppr(\nCfg + \addHapCfg - \sngConfig{\deme,\hap} + \sngConfig{\deme',\hap}) \bigg],
\end{align*}
where the normalizing constant is given by
\begin{equation*}
	\mathcal{N} = \sum_{\deme\in\demeSpace} \addHapSizeDeme{\deme} \Big[ ( \nSizeDeme{\deme}+\addHapSizeDeme{\deme} - 1 ) \relDemeSize{\deme}^{-1} + \sum_{\loc\in\locSet} \mutRate{\loc} + \sum_{\breakPoint \in \breakPointSet} \recRate{\breakPoint} + \migRateDeme{\deme} \Big].
\end{equation*}
Dividing this result by $\sampProbAppr(\nCfg)$, we thus obtain the given recursion for $\pihatmig(\addHapCfg | \nCfg)$. It is also possible to derive, in much the same way, the ``reduced'' version of this recursion; for details, see \citet{Paul2010}.

\section{Explicit transition density}
\label{app_transition_density}

We begin by assuming that the matrix $\nMatrix$ is diagonalizable, which is true if and only if $\migMatrix$ is diagonalizable. In this case, the matrix exponentials in equations~\eqref{eq_def_initial_density} and~\eqref{eq_def_transition_density} admit the eigen-decomposition $(e^{ \absTime{} \nMatrix})_{i,j} = \sum_{k=1}^{2\numDemes} e^{\absTime{} \eVal{k}}(\eVec{k} \doubleU{k})$ and $(\nMatrix e^{ \absTime{} \nMatrix})_{i,j} = \sum_{k=1}^{2\numDemes} \eVal{k} e^{\absTime{} \eVal{k}}(\eVec{k} \doubleU{k})$. Here $\eVal{k}$ are the eigenvalues of $\nMatrix$, $\eVec{k}$ are the eigenvectors, and $\doubleU{k}$ are the rows of the inverse of the matrix of eigenvectors $\vMatrix = (\eVec{1},\ldots,\eVec{2\numDemes})$. This eigen-decomposition can be used to evaluate the matrix exponential in equation~\eqref{eq_def_initial_density}, and to compute the integral in equation~\eqref{eq_def_transition_density} analytically as
\begin{equation}
	\begin{split}
		\transDensity^{(\nCfg)}_{\recRate{\breakPoint}} & (\hState{\locusIdx} \vert \hState{\locusIdx-1})  =  e^{-\recRate{\breakPoint} \absTime{\locusIdx-1}} \delta_{\hState{\locusIdx}, \hState{\locusIdx-1}}\\
			& \qquad + \frac{\nCmp{\absDeme{\locusIdx}}{\absHap{\locusIdx}}}{\nSizeDeme{\absDeme{\locusIdx}}} \frac{\recRate{\breakPoint} }{ \big( \nMatrix e^{\absTime{\locusIdx-1} \nMatrix}\big)_{\addDeme,\absState{\absDeme{\locusIdx-1}}}} \sum_{\breakDeme \in \demeSpace} \sum_{k=1}^{2\numDemes} \sum_{m=1}^{2\numDemes} \sum_{n=1}^{2\numDemes}  \Big[ ( \eVec{k} \doubleU{k})_{\addDeme,\breakDeme} ( \eVec{m} \doubleU{m})_{\breakDeme, \absState{\absDeme{\locusIdx-1}}} ( \eVec{n} \doubleU{n})_{\breakDeme, \absState{\absDeme{\locusIdx}}} \\
			& \qquad \qquad \qquad \qquad \qquad \qquad \qquad \times  \eVal{m} \eVal{n} e^{ \absTime{\locusIdx-1} \eVal{m}} e^{\absTime{\locusIdx} \eVal{n}} \integral{\eVal{k} - \eVal{m} - \eVal{n} - \recRate{\breakPoint}}{0}{\absTime{\locusIdx-1} \wedge \absTime{\locusIdx}}\Big],\\
	\end{split}
\end{equation}
where
\begin{equation}\label{eq_def_integral}
	\integral{\lambda}{a}{b} = \int_{t=a}^b e^{\lambda t} dt = \begin{cases}
				\frac{1}{\lambda} (e^{\lambda b} - e^{\lambda a}),	& \text{if $\lambda \neq 0$},\\
				b - a,	& \text{if $\lambda = 0$}.
			\end{cases}
\end{equation}
Note that for a non-diagonalizable matrix, a similar eigen-decomposition can be employed using generalized eigenvectors and the Jordan normal form, and similar, though more involved, explicit computations can be performed.    

\section{Probabilities in discretized HMM}
\label{app_discretized_probabilities}

We now give more explicit forms of the quantities involved in the probabilities of the discretized HMM, derived using the eigen-decomposition of the extended migration matrix $\nMatrix = \vMatrix \evaMatrix \vMatrix^{-1}$. Inserting equation~\eqref{eq_def_transition_density} and~\eqref{eq_def_initial_density} into the second to last line in equation~\eqref{eq_def_disc_transition_probability}, combined with the eigen-decomposition $( \nMatrix e^{ \absTime{} \nMatrix})_{i,j} = \sum_{k=1}^{2\numDemes} \eVal{k} e^{\absTime{} \eVal{k}}(\eVec{k} \doubleU{k})_{i,j}$ yields
\begin{equation}
	\noRecoTrans{\absDeme{\locusIdx-1}}{\partIdx{}}{\recRate{}} = \frac{1}{\margProb{\absDeme{}}{\partIdx{}}} \sum_{k=1}^{2\numDemes} ( \eVec{k} \doubleU{k} )_{\addDeme,\absState{\absDeme{}}}  \eVal{k} \integral{\eVal{k}-\recRate{}}{\partPoint{\partIdx{}-1}}{\partPoint{\partIdx{}}},
\end{equation}
for the first term in the last line of equation~\eqref{eq_def_disc_transition_probability}, with $\integral{\lambda}{a}{b}$ defined as in equation~\eqref{eq_def_integral}. For the second term we get
\begin{equation}
	\begin{split}
		\recoTrans{\absDeme{}}{i}{\absDeme{}'}{j}{\recRate{}}  & = \frac{\recRate{}}{\margProb{\absDeme{}}{i}} \sum_{\breakDeme \in \demeSpace} \sum_{k=1}^{2\numDemes} \sum_{m=1}^{2\numDemes} \sum_{n=1}^{2\numDemes} ( \eVec{k} \doubleU{k})_{\addDeme,\breakDeme} ( \eVec{m} \doubleU{m})_{\breakDeme, \absState{\absDeme{}}} ( \eVec{n} \doubleU{n})_{\breakDeme, \absState{\absDeme{}'}} \\
		& \qquad \qquad \qquad \times  \bigg[ e^{\eVal{m} \partPoint{i}} e^{\eVal{n} \partPoint{j}} \integral{\eVal{k} - \eVal{m} - \eVal{n} - \recRate{}}{0}{\partPoint{i} \wedge \partPoint{j}}\\
		& \qquad \qquad \qquad \qquad - e^{\eVal{m} \partPoint{i}} e^{\eVal{n} \partPoint{j-1}} \integral{\eVal{k} - \eVal{m} - \eVal{n} - \recRate{}}{0}{\partPoint{i} \wedge \partPoint{j-1}}\\
		& \qquad \qquad \qquad \qquad - e^{\eVal{m} \partPoint{i-1}} e^{\eVal{n} \partPoint{j}} \integral{\eVal{k} - \eVal{m} - \eVal{n} - \recRate{}}{0}{\partPoint{i-1} \wedge \partPoint{j}}\\
		& \qquad \qquad \qquad \qquad + e^{\eVal{m} \partPoint{i-1}} e^{\eVal{n} \partPoint{j-1}}  \integral{\eVal{k} - \eVal{m} - \eVal{n} - \recRate{}}{0}{\partPoint{i-1} \wedge \partPoint{j-1}} \bigg].\\
	\end{split}
\end{equation}
Finally, using equation~\eqref{eq_def_disc_emission_probability} one can show that
\begin{equation}
	\emDensityDisc^{(\nCfg)}_{\mutRate{}} (\addHap \vert \absDeme{},\partIdx{},\absHap{}) = \frac{1}{\margProb{\absDeme{}}{\partIdx{}}} \sum_{j=1}^{|\alleleSet{}|} \sum_{k=1}^{2\numDemes} (\eVecMut{j} \doubleUMut{j})_{\absHap{}[\locusIdx],\addHap[\locusIdx]} ( \eVec{k} \doubleU{k})_{\addDeme,\absState{\absDeme{}}} \eVal{k} \integral{\eVal{k} + \mutRate{}\eValMut{j} - \mutRate{}}{\partPoint{\partIdx{}-1}}{\partPoint{\partIdx{}}}
\end{equation}
holds, where $\alleleSet{}$ is the set of alleles at the given locus, and we used the eigen-decompositions of $\nMatrix$ and the mutation matrix $\rawMutMatrix = \vMatrixMut \,\text{diag}(\eValMut{1},\dots,\eValMut{|\alleleSet{}|})\,\vMatrixMut^{-1}$. Here $\eValMut{j}$ are the eigenvalues of the mutation matrix, $\vMatrixMut = (\eVecMut{1},\ldots,\eVecMut{2\numDemes})$ is the matrix which has the eigenvectors of the mutation matrix as columns, and $\doubleUMut{j}$ denotes the $j$-th row of $\vMatrixMut^{-1}$.

\newpage
\bibliographystyle{myplainnat}
\bibliography{bibliographyM}

\begin{thebibliography}{34}
\providecommand{\natexlab}[1]{#1}
\providecommand{\url}[1]{\texttt{#1}}
\expandafter\ifx\csname urlstyle\endcsname\relax
  \providecommand{\doi}[1]{doi: #1}\else
  \providecommand{\doi}{doi: \begingroup \urlstyle{rm}\Url}\fi

\bibitem[Cappé et~al.(2005)Cappé, Moulines, and Ryden]{Cappe2005}
Cappé, O., Moulines, E., and Ryden, T.
\newblock 2005.
\newblock \emph{Inference in Hidden Markov Models}.
\newblock Springer Series in Statistics. Springer.

\bibitem[Charlesworth(1998)]{Charlesworth1998}
Charlesworth, B.
\newblock 1998.
\newblock Measures of divergence between populations and the effect of forces
  that reduce variability.
\newblock \emph{Molecular Biology and Evolution}, {\bf 15,}\penalty0 (5)
  538--543.

\bibitem[Davison et~al.(2009)Davison, Pritchard, and Coop]{Davison2009}
Davison, D., Pritchard, J.~K., and Coop, G.
\newblock 2009.
\newblock An approximate likelihood for genetic data under a model with
  recombination and population splitting.
\newblock \emph{Theor. Popul. Biol.}, {\bf 75,}\penalty0 (4) 331--345.

\bibitem[{De~Iorio} and Griffiths(2004{\natexlab{a}})]{DeIorio2004a}
{De~Iorio}, M. and Griffiths, R.~C.
\newblock 2004{\natexlab{a}}.
\newblock Importance sampling on coalescent histories. {I}.
\newblock \emph{Adv. in Appl. Probab.}, {\bf 36,}\penalty0 (2) 417--433.

\bibitem[{De~Iorio} and Griffiths(2004{\natexlab{b}})]{DeIorio2004b}
{De~Iorio}, M. and Griffiths, R.~C.
\newblock 2004{\natexlab{b}}.
\newblock Importance sampling on coalescent histories. {II}: Subdivided
  population models.
\newblock \emph{Adv. in Appl. Probab.}, {\bf 36,}\penalty0 (2) 434--454.

\bibitem[Fearnhead and Donnelly(2001)]{Fearnhead2001}
Fearnhead, P. and Donnelly, P.
\newblock 2001.
\newblock Estimating recombination rates from population genetic data.
\newblock \emph{Genetics}, {\bf 159,} 1299--1318.

\bibitem[Gay et~al.(2007)Gay, Myers, and McVean]{Gay2007}
Gay, J., Myers, S.~R., and McVean, G. A.~T.
\newblock 2007.
\newblock Estimating meiotic gene conversion rates from population genetic
  data.
\newblock \emph{Genetics}, {\bf 177,} 881--894.

\bibitem[Gravel et~al.(2011)Gravel, Henn, Gutenkunst, Indap, Marth, Clark, Yu,
  Gibbs, Project, and Bustamante]{Gravel2011}
Gravel, S., Henn, B.~M., Gutenkunst, R.~N., Indap, A.~R., Marth, G.~T., Clark,
  A.~G., Yu, F., Gibbs, R.~A., Project, T. .~G., and Bustamante, C.~D.
\newblock 2011.
\newblock Demographic history and rare allele sharing among human populations.
\newblock \emph{Proceedings of the National Academy of Sciences}.

\bibitem[Griffiths and Marjoram(1997)]{Griffiths1997}
Griffiths, R.~C. and Marjoram, P.
\newblock 1997.
\newblock An ancestral recombination graph.
\newblock In \emph{Progress in population genetics and human evolution
  (Minneapolis, MN, 1994)}, volume~87 of \emph{IMA Vol. Math. Appl.}, pages
  257--270. Springer, New York.

\bibitem[Griffiths et~al.(2008)Griffiths, Jenkins, and Song]{Griffiths2008}
Griffiths, R.~C., Jenkins, P.~A., and Song, Y.~S.
\newblock 2008.
\newblock Importance sampling and the two-locus model with subdivided
  population structure.
\newblock \emph{Adv. in Appl. Probab.}, {\bf 40,}\penalty0 (2) 473--500.

\bibitem[Gutenkunst et~al.(2009)Gutenkunst, Hernandez, Williamson, and
  Bustamante]{Gutenkunst2009}
Gutenkunst, R.~N., Hernandez, R.~D., Williamson, S.~H., and Bustamante, C.~D.
\newblock 10 2009.
\newblock Inferring the joint demographic history of multiple populations from
  multidimensional {SNP} frequency data.
\newblock \emph{PLoS Genet}, {\bf 5,}\penalty0 (10) e1000695.

\bibitem[Hellenthal et~al.(2008)Hellenthal, Auton, and Falush]{Hellenthal2008}
Hellenthal, G., Auton, A., and Falush, D.
\newblock 2008.
\newblock Inferring human colonization history using a copying model.
\newblock \emph{PLoS Genet.}, {\bf 4,}\penalty0 (5) e1000078.

\bibitem[Herbots(1997)]{Herbots1997}
Herbots, H.~M.
\newblock 1997.
\newblock The structured coalescent.
\newblock In \emph{Progress in population genetics and human evolution
  (Minneapolis, MN, 1994)}, volume~87 of \emph{IMA Vol. Math. Appl.}, pages
  231--255. Springer, New York.

\bibitem[Howie et~al.(2009)Howie, Donnelly, and Marchini]{Howie2009}
Howie, B.~N., Donnelly, P., and Marchini, J.
\newblock 2009.
\newblock A flexible and accurate genotype imputation method for the next
  generation of genome-wide association studies.
\newblock \emph{PLoS Genet}, {\bf 5,}\penalty0 (6) e1000529.

\bibitem[Lawson et~al.(2012)Lawson, Hellenthal, Myers, and Falush]{Lawson2012}
Lawson, D., Hellenthal, G., Myers, S., and Falush, D.
\newblock 2012.
\newblock Inference of population structure using dense haplotype data.
\newblock \emph{PLoS Genetics}, {\bf 8,}\penalty0 (1) e1002453.

\bibitem[Li and Durbin(2011)]{Li2011}
Li, H. and Durbin, R.
\newblock 2011.
\newblock Inference of human population history from individual whole-genome
  sequences.
\newblock \emph{Nature}, {\bf 475,} 493--496.

\bibitem[Li and Stephens(2003)]{Li2003}
Li, N. and Stephens, M.
\newblock 2003.
\newblock Modelling linkage disequilibrium, and identifying recombination
  hotspots using {SNP} data.
\newblock \emph{Genetics}, {\bf 165,} 2213--2233.

\bibitem[Li and Abecasis(2006)]{Li2006}
Li, Y. and Abecasis, G.~R.
\newblock 2006.
\newblock Mach 1.0: Rapid haplotype reconstruction and missing genotype
  inference.
\newblock \emph{Am. J. Hum. Genet.}, {\bf S79,} 2290.

\bibitem[Li et~al.(2010)Li, Willer, Ding, Scheet, and Abecasis]{Li2010}
Li, Y., Willer, C.~J., Ding, J., Scheet, P., and Abecasis, G.~R.
\newblock 2010.
\newblock Mach: Using sequence and genotype data to estimate haplotypes and
  unobserved genotypes.
\newblock \emph{Genetic Epidemiology}, {\bf 34,} 816--834.

\bibitem[Mailund et~al.(2011)Mailund, Dutheil, Hobolth, Lunter, and
  Schierup]{Mailund2011}
Mailund, T., Dutheil, J.~Y., Hobolth, A., Lunter, G., and Schierup, M.~H.
\newblock 03 2011.
\newblock Estimating divergence time and ancestral effective population size of
  bornean and sumatran orangutan subspecies using a coalescent hidden markov
  model.
\newblock \emph{PLoS Genet}, {\bf 7,}\penalty0 (3) e1001319.

\bibitem[Marchini et~al.(2007)Marchini, Howie, Myers, McVean, and
  Donnelly]{Marchini2007}
Marchini, J., Howie, B., Myers, S.~R., McVean, G. A.~T., and Donnelly, P.
\newblock 2007.
\newblock A new multipoint method for genome-wide association studies by
  imputation of genotypes.
\newblock \emph{Nat. Genet.}, {\bf 39,}\penalty0 (7) 906--13.

\bibitem[Marjoram and Wall(2006)]{Marjoram2006}
Marjoram, P. and Wall, J.~D.
\newblock 2006.
\newblock Fast ``coalescent'' simulation.
\newblock \emph{BMC Genet.}, {\bf 7,} 16.

\bibitem[McVean and Cardin(2005)]{McVean2005}
McVean, G.~A. and Cardin, N.~J.
\newblock 2005.
\newblock Approximating the coalescent with recombination.
\newblock \emph{Philos. Trans. R. Soc. Lond. B Biol. Sci.}, {\bf 360,}
  1387--93.

\bibitem[Myers et~al.(2008)Myers, Fefferman, and Patterson]{Myers2008}
Myers, S., Fefferman, C., and Patterson, N.
\newblock 2008.
\newblock Can one learn history from the allelic spectrum?
\newblock \emph{Theoretical Population Biology}, {\bf 73,}\penalty0 (3)
  342--348.

\bibitem[Paul and Song(2010)]{Paul2010}
Paul, J.~S. and Song, Y.~S.
\newblock 2010.
\newblock A principled approach to deriving approximate conditional sampling
  distributions in population genetics models with recombination.
\newblock \emph{Genetics}, {\bf 186,} 321--338.

\bibitem[Paul and Song(2012)]{Paul2012}
Paul, J.~S. and Song, Y.~S.
\newblock 2012.
\newblock Blockwise {HMM} computation for large-scale population genomic
  inference.
\newblock \emph{Bioinformatics (submitted)}.

\bibitem[Paul et~al.(2011)Paul, Steinr\"ucken, and Song]{Paul2011}
Paul, J.~S., Steinr\"ucken, M., and Song, Y.~S.
\newblock 2011.
\newblock An accurate sequentially {M}arkov conditional sampling distribution
  for the coalescent with recombination.
\newblock \emph{Genetics}, {\bf 187,} 1115--1128.

\bibitem[Price et~al.(2009)Price, Tandon, Patterson, Barnes, Rafaels,
  Ruczinski, Beaty, Mathias, Reich, and Myers]{Price2009}
Price, A.~L., Tandon, A., Patterson, N., Barnes, K.~C., Rafaels, N., Ruczinski,
  I., Beaty, T.~H., Mathias, R., Reich, D., and Myers, S.
\newblock 2009.
\newblock Sensitive detection of chromosomal segments of distinct ancestry in
  admixed populations.
\newblock \emph{PLoS Genet.}, {\bf 5,}\penalty0 (6) e1000519.

\bibitem[Stephens and Donnelly(2000)]{Stephens2000}
Stephens, M. and Donnelly, P.
\newblock 2000.
\newblock Inference in molecular population genetics.
\newblock \emph{J. R. Stat. Soc. Ser. B Stat. Methodol.}, {\bf 62,}\penalty0
  (4) 605--655.

\bibitem[Stephens and Scheet(2005)]{Stephens2005}
Stephens, M. and Scheet, P.
\newblock 2005.
\newblock Accounting for decay of linkage disequilibrium in haplotype inference
  and missing-data imputation.
\newblock \emph{Am. J. Hum. Genet.}, {\bf 76,}\penalty0 (3) 449--462.

\bibitem[Wang and Hey(2010)]{Wang2010}
Wang, Y. and Hey, J.
\newblock 2010.
\newblock Estimating divergence parameters with small samples from a large
  number of loci.
\newblock \emph{Genetics}, {\bf 184,}\penalty0 (2) 363--379.

\bibitem[Wegmann et~al.(2011)Wegmann, Kessner, Veeramah, Mathias, Nicolae,
  Yanek, Sun, Torgerson, Rafaels, Mosley, Becker, Ruczinski, Beaty, Kardia,
  Meyers, Barnes, Becker, Freimer, and Novembre]{Wegmann2011}
Wegmann, D., Kessner, D.~E., Veeramah, K.~R., Mathias, R.~A., Nicolae, D.~L.,
  Yanek, L.~R., Sun, Y.~V., Torgerson, D.~G., Rafaels, N., Mosley, T., Becker,
  L.~C., Ruczinski, I., Beaty, T.~H., Kardia, S. L.~R., Meyers, D.~A., Barnes,
  K.~C., Becker, D.~M., Freimer, N.~B., and Novembre, J.
\newblock 2011.
\newblock Recombination rates in admixed individuals identified by
  ancestry-based inference.
\newblock \emph{Nat. Genet.}, {\bf 43,} 847--853.

\bibitem[Wiuf and Hein(1999)]{Wiuf1999}
Wiuf, C. and Hein, J.
\newblock 1999.
\newblock Recombination as a point process along sequences.
\newblock \emph{Theor. Pop. Biol.}, {\bf 55,} 248--259.

\bibitem[Yin et~al.(2009)Yin, Jordan, and Song]{Yin2009}
Yin, J., Jordan, M.~I., and Song, Y.~S.
\newblock 2009.
\newblock Joint estimation of gene conversion rates and mean conversion tract
  lengths from population {SNP} data.
\newblock \emph{Bioinformatics}, {\bf 25,}\penalty0 (12) i231--9.

\end{thebibliography}

\end{document}